\documentclass[final, a4paper, 10pt,twocolumn]{IEEEtran}

\usepackage{import,bm,url,textcomp,graphicx,cite,balance}
\usepackage{amsmath,amssymb,amsfonts,mathtools}

\usepackage[T1]{fontenc}
\usepackage[latin1]{inputenc}
\usepackage[table]{xcolor}

\usepackage{hyperref}
\hypersetup{colorlinks=true,linkcolor=black,urlcolor=black,}

\usepackage{multirow}
\usepackage[export]{adjustbox}
\usepackage{arydshln} 
\usepackage{pifont}
\usepackage{flushend}
\newcommand{\cmark}{\ding{52}}  

\usepackage{epstopdf}

\usepackage{enumitem}

\usepackage{cuted}
\usepackage{needspace}  
\newenvironment{bottomborder}%
{}  
{\needspace{\baselineskip}\hrule} 

\usepackage{dirtytalk}

\usepackage{anyfontsize}

\usepackage{caption}
\usepackage[caption=false]{subfig} 
\makeatletter
\renewcommand{\fnum@figure}{Fig. \thefigure}
\makeatother

\usepackage[acronym,toc,shortcuts]{glossaries}
\newacronym{Ph.D.}{Ph.D.}{Doctor of Philosophy}
\newacronym{SA}{SA}{Simulated Annealing}
\newacronym{VLC}{VLC}{visible light communications}
\newacronym{RF}{RF}{radio frequency}
\newacronym{V2V}{V2V}{vehicle-to-vehicle}
\newacronym{V2X}{V2X}{vehicle-to-everything}
\newacronym{V2I}{V2I}{vehicle-to-infrastructure}
\newacronym{B5G}{B5G}{beyond-fifth generation}
\newacronym{LED}{LED}{light-emitting diode}
\newacronym{OMA}{OMA}{orthogonal multiple access}
\newacronym{FDMA}{FDMA}{frequency-division multiple-access}
\newacronym{TDMA}{TDMA}{time-division multiple-access}
\newacronym{CDMA}{CDMA}{code-division multiple-access}
\newacronym{OFDMA}{OFDMA}{orthogonal frequency-division multiple-access}
\newacronym{OFDM}{OFDM}{orthogonal frequency-division multiplexing}
\newacronym{WDMA}{WDMA}{wavelength-division multiple-access}
\newacronym{NOMA}{NOMA}{non-orthogonal multiple access}
\newacronym{PD-NOMA}{PD-NOMA}{power-domain NOMA}
\newacronym{CD-NOMA}{CD-NOMA}{code-domain NOMA}
\newacronym{SC}{SC}{superposition coding}
\newacronym{SIC}{SIC}{successive interference cancellation}
\newacronym{BS}{BS}{base station}
\newacronym{QoS}{QoS}{quality-of-service}
\newacronym{NP}{NP}{non-deterministic polynomial-time}
\newacronym{DCO-OFDM}{DCO-OFDM}{direct-current biased optical-OFDM}
\newacronym{DCO-OFDMA}{DCO-OFDMA}{direct-current biased optical-OFDMA}
\newacronym{DC}{DC}{direct current}
\newacronym{ITU}{ITU}{international telecommunication union}
\newacronym{FoV}{FoV}{field-of-view}
\newacronym{CSI}{CSI}{channel state information}
\newacronym{LACO-OFDM}{LACO-OFDM}{layered asymmetrically clipped optical OFDM}
\newacronym{ACO-OFDM}{ACO-OFDM}{asymmetrically clipped optical OFDM}
\newacronym{FR}{FR}{frequency reuse}
\newacronym{EAs}{EAs}{evolutionary algorithms}
\newacronym{C-LiAN}{C-LiAN}{centralized light access network}
\newacronym{AP}{AP}{access point}
\newacronym{PD}{PD}{photo-diode}
\newacronym{SINR}{SINR}{signal-to-noise-interference ratio}
\newacronym{LoS}{LoS}{line-of-sight}
\newacronym{AWGN}{AWGN}{additive white Gaussian noise}
\newacronym{SNR}{SNR}{signal-to-noise ratio}
\newacronym{NLUPA}{NLUPA}{next-largest-difference user-pairing algorithm}
\newacronym{D-NLUPA}{D-NLUPA}{divide-and-next-largest-difference user-pairing algorithm}
\newacronym{NAICS}{NAICS}{network-assisted interference cancellation and suppression}
\newacronym{LTE}{LTE}{long term evolution}
\newacronym{3GPP}{3GPP}{3rd Generation Partnership Project}
\newacronym{CR}{CR}{cognitive radio}
\newacronym{2D}{2D}{two-dimension}
\newacronym{GP}{GP}{gradient projection}
\newacronym{umMTC}{umMTC}{ultra-massive machine-type communication}
\newacronym{mMTC}{mMTC}{massive machine-type communication}
\newacronym{IoE}{IoE}{internet-of-everything}
\newacronym{IoUT}{IoUT}{internet-of-underwater-things}
\newacronym{ZF}{ZF}{zero-forcing}
\newacronym{NLIP}{NLIP}{non-linear integer programming}
\newacronym{DP}{DP}{dynamic programming}
\newacronym{MIMO}{MIMO}{multiple-input multiple-output}
\newacronym{TS}{TS}{Tabu-search}
\newacronym{THz}{THz}{Terahertz}
\newacronym{MISO}{MISO}{multiple-input single-output}
\newacronym{SIMO}{SIMO}{single-input multiple-output}
\newacronym{EE}{EE}{energy efficiency}
\newacronym{VR}{VR}{virtual reality}
\newacronym{XR}{XR}{extended reality}
\newacronym{5G}{5G}{fifth generation}
\newacronym{6G}{6G}{sixth generation}
\newacronym{NR}{NR}{new radio}
\newacronym{mmWave}{mmWave}{millimeter-wave}
\newacronym{FD}{FD}{full-duplex}
\newacronym{CNOMA}{CNOMA}{cooperative NOMA}
\newacronym{ABF}{ABF}{analog beamforming}
\newacronym{BF}{BF}{beamforming}
\newacronym{DF}{DF}{decode-and-forward}
\newacronym{AF}{AF}{amplify-and-forward}
\newacronym{CF}{CF}{compress-and-forward}
\newacronym{SPS}{SPS}{single-phase shifter}
\newacronym{PS}{PS}{phase shifter}
\newacronym{PA}{PA}{power amplifier}
\newacronym{NLoS}{NLoS}{non-line-of-sight}
\newacronym{ULA}{ULA}{uniform linear array}
\newacronym{SI}{SI}{self-interference}
\newacronym{MA}{MA}{multiple access}
\newacronym{1G}{1G}{first generation}
\newacronym{2G}{2G}{second generation}
\newacronym{3G}{3G}{third generation}
\newacronym{4G}{4G}{fourth generation}
\newacronym{M2M}{M2M}{machine-to-machine}
\newacronym{IoT}{IoT}{internet-of-things}
\newacronym{IMT}{IMT}{International Mobile Telecommunications}
\newacronym{SE}{SE}{spectral efficiency}
\newacronym{MMF}{MMF}{Maximin Fairness}
\newacronym{WSR}{WSR}{weighted sum rate}
\newacronym{D-NOMA}{D-NOMA}{dynamic-NOMA}
\newacronym{GSM}{GSM}{global system for mobile telecommunications}
\newacronym{IS-95}{IS-95}{Interim Standard 95}
\newacronym{SMS}{SMS}{short-message service}
\newacronym{CoMP}{CoMP}{coordinated multi-point}
\newacronym{MU-MIMO}{MU-MIMO}{multi-user MIMO}
\newacronym{MAC}{MAC}{multiple-access channel}
\newacronym{BC}{BC}{vector-broadcast channel}
\newacronym{CSIT}{CSIT}{channel state information at the transmitter}
\newacronym{DPC}{DPC}{dirty paper coding}
\newacronym{ZF-DPC}{ZF-DPC}{zero-forcing DPC}
\newacronym{BD}{BD}{block-diagonalization}
\newacronym{ICI}{ICI}{inter-cell interference}
\newacronym{MMSE}{MMSE}{minimum mean square error}
\newacronym{LTE-A}{LTE-A}{long term evolution-advanced}
\newacronym{MUST}{MUST}{multi-user superposition transmission}
\newacronym{SISO}{SISO}{single-input single-output}
\newacronym{LDS-CDMA}{LDS-CDMA}{low-density spreading CDMA}
\newacronym{LDS-OFDM}{LDS-OFDM}{low-density spreading OFDM}
\newacronym{SCMA}{SCMA}{sparse code multiple access}
\newacronym{MUSA}{MUSA}{multi-user sharing access}
\newacronym{SAMA}{SAMA}{successive interference cancellation amenable multiple access}
\newacronym{PDMA}{PDMA}{pattern division multiple access} 
\newacronym{BOMA}{BOMA}{building block sparse-constellation based orthogonal multiple access}  
\newacronym{LPMA}{LPMA}{lattice partition multiple access} \newacronym{OOK}{OOK}{on-off keying} 
\newacronym{M-PAM}{M-PAM}{M-ary pulse-amplitude modulation} 
\newacronym{M-PPM}{M-PPM}{M-ary pulse-position modulation} \newacronym{MSM}{MSM}{multiple-subcarrier modulation}
\newacronym{IM/DD}{IM/DD}{intensity modulation and direct detection}
\newacronym{RC}{RC}{repetition code}
\newacronym{SM}{SM}{spatial multiplexing}
\newacronym{SMOD}{SMOD}{spatial modulation}
\newacronym{BER}{BER}{bit error rate}
\newacronym{SER}{SER}{symbol error rate}
\newacronym{MFTP}{MFTP}{maximum flickering time period}
\newacronym{MU-MISO}{MU-MISO}{multi-user multi-input single-output}
\newacronym{MSE}{MSE}{minimum square error}
\newacronym{SPCA}{SPCA}{sequential parametric convex approximation}
\newacronym{WSMSE}{WSMSE}{weighted sum minimum square error}
\newacronym{OCDMA}{OCDMA}{optical code division multiple access}
\newacronym{SDMA}{SDMA}{space-division multiple access}
\newacronym{DMT}{DMT}{discrete multi-tone}
\newacronym{VLNs}{VLNs}{visible light networks}
\newacronym{VHO}{VHO}{Vertical handover}
\newacronym{RSS}{RSS}{received signal strength}
\newacronym{RSI}{RSI}{received signal intensity}
\newacronym{IA}{IA}{interference alignment}
\newacronym{BIA}{BIA}{blind interference alignment}
\newacronym{BBU}{BBU}{base-band unit}
\newacronym{SU}{SU}{secondary user}
\newacronym{PU}{PU}{primary user}
\newacronym{mMIMO}{mMIMO}{massive-MIMO}
\newacronym{UAV}{UAV}{unmanned aerial vehicle}
\newacronym{PHY}{PHY}{physical}
\newacronym{CF-mMIMO}{CF-mMIMO}{cell-free mMIMO}
\newacronym{LIS}{LIS}{large intelligent surfaces}
\newacronym{3-D MIMO}{3-D MIMO}{3-Dimensional MIMO}
\newacronym{RIS}{RIS}{reflecting intelligent surface}
\newacronym{BackCom}{BackCom}{backscatter communications}
\newacronym{UL}{UL}{uplink}
\newacronym{UE}{UE}{user equipment}
\newacronym{D2D}{D2D}{device-to-device}
\newacronym{FCC}{FCC}{Federal Communications Commission}
\newacronym{HAP}{HAP}{high altitude platform}
\newacronym{LAP}{LAP}{low altitude platform}
\newacronym{MEC}{MEC}{mobile edge computing}
\newacronym{NATO}{NATO}{North Atlantic Treaty Organization}
\newacronym{ML}{ML}{machine learning}
\newacronym{QML}{QML}{quantum machine learning}
\newacronym{DL}{DL}{deep learning}
\newacronym{DRL}{DRL}{deep reinforcement learning}
\newacronym{RL}{RL}{Reinforcement learning}
\newacronym{MMA}{MMA}{minorization maximization algorithm}
\newacronym{MM}{MM}{majorization-minimization}
\newacronym{KKT}{KKT}{Karush–Kuhn–Tucker}
\newacronym{FDD}{FDD}{frequency division duplex}
\newacronym{SCA}{SCA}{sine-cosine algorithm}
\newacronym{AoD}{AoD}{angle-of-departure}
\newacronym{SDP}{SDP}{semi-definite programming}
\newacronym{SDR}{SDR}{semi-definite relaxation}
\newacronym{JT}{JT}{joint transmission}
\newacronym{CB}{CB}{coordinated beamforming}
\newacronym{RAMA}{RAMA}{relay-aided multiple access}
\newacronym{MRC}{MRC}{maximum ratio combining}
\newacronym{SWIPT}{SWIPT}{simultaneous wireless information and power transfer}
\newacronym{HetNets}{HetNets}{heterogeneous networks}
\newacronym{D.C.}{D.C.}{difference of convex}
\newacronym{GRPA}{GRPA}{gain ratio power allocation}
\newacronym{FPA}{FPA}{fixed power allocation}
\newacronym{CS}{CS}{Cuckoo Search}
\newacronym{HHO}{HHO}{Harris Hawks Optimizer}
\newacronym{PLC}{PLC}{power line communications}
\newacronym{HTT}{HTT}{harvest-then-transmit}
\newacronym{H-CRAN}{H-CRAN}{heterogeneous cloud radio access network}
\newacronym{RRHs}{RRHs}{remote radio heads}
\newacronym{IIoT}{IIoT}{Industrial IoT}
\newacronym{PAPR}{PAPR}{peak-to-average-power-ratio}
\newacronym{ANC}{ANC}{analog network coding}
\newacronym{FFR}{FFR}{fractional frequency reuse}
\newacronym{RGB}{RGB}{red-green-blue}
\newacronym{MAR}{MAR}{mobile augmented reality}
\newacronym{HD}{HD}{half-duplex}
\newacronym{CPU}{CPU}{central process unit}
\newacronym{RWP}{RWP}{Random Way-Point}
\newacronym{LC}{LC}{liquid crystal}
\newacronym{ADR}{ADR}{angle diversity receiver}
\newacronym{RSMA}{RSMA}{rate splitting multiple access}
\newacronym{omni-DRIS}{omni-DRIS}{omni-digital-RIS}
\newacronym{STAR-RIS}{STAR-RIS}{simultaneous transmission and reflection reconfigurable intelligent surface}
\newacronym{OSTAR-RIS}{OSTAR-RIS}{optical simultaneous transmission and reflection reconfigurable intelligent surface}
\newacronym{AN}{AN}{artificial noise}
\newacronym{PLS}{PLS}{physical layer security}
\newacronym{AR}{AR}{augmented reality}
\newacronym{HRS}{HRS}{hierarchical RS}
\newacronym{GRS}{GRS}{generalized RS}
\newacronym{RS}{RS}{rate splitting}
\newacronym{DDPG}{DDPG}{deep deterministic policy gradient}
\newacronym{ORIS}{ORIS}{optical RIS}
\newacronym{SR}{SR}{secrecy rate}
\newacronym{SSR}{SSR}{sum secrecy rate}
\newacronym{SEE}{SEE}{secrecy energy efficiency}
\newacronym{MINLP}{MINLP}{mixed-integer non-linear programming}
\newacronym{GA}{GA}{genetic algorithm}
\glsdisablehyper

\usepackage[linesnumbered,ruled,vlined]{algorithm2e}

\begin{document}

\bstctlcite{IEEEexample:BSTcontrol} 

\title{Max-Min Secrecy Rate and Secrecy Energy Efficiency Optimization for RIS-Aided VLC Systems: RSMA Versus NOMA}

\author{Omar~Maraqa, Sylvester~Aboagye,~\IEEEmembership{Member,~IEEE}, Majid~H.~Khoshafa,~\IEEEmembership{Senior Member,~IEEE},  and~Telex~M.~N.~Ngatched,~\IEEEmembership{Senior Member,~IEEE}%

\thanks{
O. Maraqa, M. H. Khoshafa, and T. M. N. Ngatched are with the Department of Electrical and Computer Engineering, McMaster University, Hamilton, ON L8S 4L8, Canada (e-mail:\{maraqao@mcmaster.ca, khoshafm@mcmaster.ca, and ngatchet@mcmaster.ca\}).

S. Aboagye is with the School of Engineering, University of Guelph, Guelph, ON N1G 2W1, Canada (e-mail: \{saboagye@uoguelph.ca\}).

} }

\markboth{Accepted by IEEE Open Journal of Vehicular Technology, May 2025}%
{Maraqa~\MakeLowercase{\textit{et al.}}: Max-Min Secrecy Rate and Secrecy Energy Efficiency Optimization for RIS-Aided VLC Systems: RSMA Versus NOMA}

\maketitle

\begin{abstract}
Integrating visible light communication (VLC) with the reconfigurable intelligent surface (RIS) significantly enhances physical layer security by enabling precise directional signal control and dynamic adaptation to the communication environment. These capabilities strengthen the confidentiality and security of VLC systems. This paper presents a comprehensive study on the joint optimization of VLC access point (AP) power allocation, RIS association, and RIS elements orientation angles for secure VLC systems, while considering rate-splitting multiple access (RSMA) and power-domain non-orthogonal multiple access (NOMA) schemes. Specifically, two frameworks are proposed to maximize both the minimum secrecy rate (SR) and the minimum secrecy energy efficiency (SEE) by jointly optimizing power allocation, RIS association, and RIS elements orientation angles for both power-domain NOMA and RSMA-based VLC systems. The proposed frameworks consider random device orientation and guarantee the minimum user-rate requirement. The proposed optimization frameworks belong to the class of mixed integer nonlinear programming, which has no known feasible solution methodology to guarantee the optimal solution. Moreover, the increased degree of freedom and flexibility from the joint consideration of power control, RIS association and element orientation results in a large set of decision variables and constraints, which further complicates the optimization problem. To that end, we utilize a genetic algorithm-based solution method, which through its exploration and exploitation capabilities can obtain a good quality solution. Additionally, comprehensive simulations show that the RSMA scheme outperforms the power-domain NOMA scheme across both the SR and SEE metrics over various network parameters. Furthermore, useful insights on the impact of minimum user rate requirement, number of RIS elements, and maximum VLC AP transmit power on the minimum SR and SEE performances are provided.
\end{abstract}


\section{Introduction}
\Gls{VLC} is increasingly recognized as a vital technology for next-generation wireless communications, especially within \gls{6G} networks. In \gls{VLC}, the visible light spectrum ($400$-$700$ nm) is utilized for data transmission, offering several advantages over conventional \gls{RF} communications~\cite{7239528}. For instance, \gls{VLC} provides a vastly larger unlicensed bandwidth, alleviating spectrum scarcity and supporting high data rates necessary for \gls{6G} applications like \gls{VR} and \gls{AR}~\cite{9208801}. Moreover, \gls{VLC} systems are immune to electromagnetic interference, making them suitable for environments where RF interference is prevalent or RF communication is restricted. However, \gls{VLC} faces challenges such as the \gls{LoS} requirement, limited coverage area, and susceptibility to security threats due to its broadcasting nature. Some solutions, including advanced beam steering technologies, multi-hop communication strategies, and dense networks of VLC transmitters, were proposed to extend coverage and improve signal quality~\cite{8698841,8528460}. Nonetheless, more research is needed to pave the way for \gls{VLC}'s successful and secure integration into \gls{6G} networks. Towards this end, one possible solution is to integrate \glspl{RIS} into \gls{VLC} systems to achieve \gls{PLS}.

\Glspl{RIS} represents a cutting-edge solution for wireless communication, operating as a controllable mirror array that dynamically steers light toward a desired direction. By employing light reflection and refraction properties, \Glspl{RIS} selectively enhances signal strength at intended receivers while minimizing interference at unintended locations, resulting in improved signal propagation and transmission quality in real-time according to user requirements~\cite{10736549}. This adaptability makes \glspl{RIS} ideal for \gls{VLC}, improving signal quality, enhancing security, extending coverage, increasing data rates, and enhancing energy efficiency~\cite{9968053}. Transmission strategies for vehicle-to-everything communications enhanced by RISs were explored in~\cite{9744412}. Using RISs in vehicle platoons was investigated in~\cite{10158691}, with coordination from a \gls{BS} to improve high-precision location tracking. Channel tracking methods for RIS-assisted high-mobility mmWave communication systems were explored in~\cite{9984171}.

To facilitate ubiquitous multi-user communications in \gls{VLC} systems, non-orthogonal multiple access techniques such as \gls{RSMA} and power-domain \gls{NOMA} have been integrated into \gls{VLC} systems~\cite{9968053}. \Gls{RSMA} has emerged as a promising transmission technique employed in multi-antenna wireless networks that improves network performance by dividing each user's message into common and private sub-messages. These sub-messages are then decoded exclusively by their respective users, enabling flexible interference management. \Gls{RSMA} also improves efficiency by partially decoding interference and treating the remaining interference as noise~\cite{9831440,10038476}. Towards this end, \gls{RSMA} enhances \gls{PLS} in wireless communication networks through improved interference management, adaptive message splitting, and flexible transmission models~\cite{10155552}. By partially decoding interference and treating the rest as noise, \gls{RSMA} ensures high-quality signal reception for legitimate users, reducing susceptibility to eavesdropping and passive attacks. \gls{RSMA} dynamic partitioning of messages into common and private messages provides more secure transmission. Separating these messages enhances confidentiality, increases secrecy capacity, and efficiently uses spectral and power resources~\cite{9585491}. These attributes make \gls{RSMA} a robust solution for securing modern wireless communication systems against sophisticated threats. On the other hand, given the potential of \gls{PLS} in future networks, developing \gls{PLS} techniques for power-domain \gls{NOMA} and exploring associated security challenges are compelling research areas. Although the power-domain \gls{NOMA}'s major purpose is to facilitate the concurrent transmission of information to multiple users over shared radio resources~\cite{9154358}, its distinct features can be leveraged to address security vulnerabilities and enhance the confidentiality and integrity of wireless communications. By utilizing the power-domain \gls{NOMA}'s key capabilities, such as \gls{SIC} and power allocation, innovative security strategies can be designed to prevent eavesdropping and improve the overall security performance of wireless networks~\cite{10185552, 10598216}.

\begin{table*}[!t]
\centering
\caption{State-of-the-art \gls{RIS}-aided secure \gls{VLC} systems. (``SR'':``Secrecy rate'', ``SSR'':``Sum secrecy rate'', and ``SEE'':``Secrecy energy efficiency'')}
\label{Table_I}
\resizebox{\textwidth}{!}{%
\begin{tabular}{|c|ccc|cccc|ccc|c|}
\hline
\multirow{2}{*}{\textbf{{[}\#{]}}} & \multicolumn{3}{c|}{\textbf{Optimization decision variables}} & \multicolumn{4}{c|}{\textbf{Optimized objective function}} & \multicolumn{3}{c|}{\textbf{Access technique}} & \multicolumn{1}{c|}{\textbf{Rx configuration}} \\ \cline{2-12} 

 & \multicolumn{1}{c|}{\begin{tabular}[c]{@{}c@{}} \textbf{RIS} \\ \textbf{elements} \\ \textbf{orientation} \\ \textbf{angles} \end{tabular}} & \multicolumn{1}{c|}{\begin{tabular}[c]{@{}c@{}} \textbf{RIS-user} \\ \textbf{association} \end{tabular}} & \multicolumn{1}{c|}{\begin{tabular}[c]{@{}c@{}} \textbf{VLC AP} \\ \textbf{power} \\ \textbf{allocation}  \end{tabular} } & \multicolumn{1}{c|}{\textbf{Max. SR}} & \multicolumn{1}{c|}{\textbf{Max. SSR}} & \multicolumn{1}{c|}{\textbf{Max-Min SR}} & \textbf{Max-Min SEE} & \multicolumn{1}{c|}{\textbf{OMA}} & \multicolumn{1}{c|}{\textbf{NOMA}} & \textbf{RSMA} & \multicolumn{1}{c|}{\textbf{\begin{tabular}[c]{@{}c@{}} Random \\device \\ orientation \end{tabular} }} \\ \hline

\cite{9756553}  & \multicolumn{1}{c|}{} & \multicolumn{1}{c|}{\cmark} & \multicolumn{1}{c|}{} & \multicolumn{1}{c|}{} & \multicolumn{1}{c|}{\cmark} & \multicolumn{1}{c|}{} &  & \multicolumn{1}{c|}{\cmark} & \multicolumn{1}{c|}{} &  & \\ \hline

\cite{9784887}  & \multicolumn{1}{c|}{\cmark} & \multicolumn{1}{c|}{} & \multicolumn{1}{c|}{} & \multicolumn{1}{c|}{\cmark} & \multicolumn{1}{c|}{} & \multicolumn{1}{c|}{} &  & \multicolumn{1}{c|}{\cmark} & \multicolumn{1}{c|}{} &  & \\ \hline

\cite{10511290}  & \multicolumn{1}{c|}{\cmark} & \multicolumn{1}{c|}{} & \multicolumn{1}{c|}{\cmark} & \multicolumn{1}{c|}{\cmark} & \multicolumn{1}{c|}{} & \multicolumn{1}{c|}{} & \multicolumn{1}{c|}{} & \multicolumn{1}{c|}{\cmark} & \multicolumn{1}{c|}{} &  & \\ \hline

\cite{10516681}  & \multicolumn{1}{c|}{} & \multicolumn{1}{c|}{\cmark} & \multicolumn{1}{c|}{} & \multicolumn{1}{c|}{\cmark} & \multicolumn{1}{c|}{} & \multicolumn{1}{c|}{} & \multicolumn{1}{c|}{} & \multicolumn{1}{c|}{\cmark} &\multicolumn{1}{c|}{} &  & \\ \hline

\cite{9500409}  & \multicolumn{1}{c|}{\cmark} & \multicolumn{1}{c|}{} & \multicolumn{1}{c|}{} &\multicolumn{1}{c|}{\cmark} & \multicolumn{1}{c|}{} & \multicolumn{1}{c|}{} &  & \multicolumn{1}{c|}{\cmark} & \multicolumn{1}{c|}{} &  & \\ \hline

\cite{10669590}  & \multicolumn{1}{c|}{} & \multicolumn{1}{c|}{\cmark} & \multicolumn{1}{c|}{} &  \multicolumn{1}{c|}{\cmark} & \multicolumn{1}{c|}{} & \multicolumn{1}{c|}{} & \multicolumn{1}{c|}{} & \multicolumn{1}{c|}{\cmark} &\multicolumn{1}{c|}{} &  & \\ \hline

\cite{10279487}  & \multicolumn{1}{c|}{} & \multicolumn{1}{c|}{\cmark} & \multicolumn{1}{c|}{\cmark} &  \multicolumn{1}{c|}{\cmark} & \multicolumn{1}{c|}{} & \multicolumn{1}{c|}{} & \multicolumn{1}{c|}{} & \multicolumn{1}{c|}{} &\multicolumn{1}{c|}{\cmark} &  & \\ \hline

\cite{10746540}  & \multicolumn{1}{c|}{} & \multicolumn{1}{c|}{\cmark} & \multicolumn{1}{c|}{\cmark} &  \multicolumn{1}{c|}{} & \multicolumn{1}{c|}{} & \multicolumn{1}{c|}{\cmark} & \multicolumn{1}{c|}{} & \multicolumn{1}{c|}{} &\multicolumn{1}{c|}{\cmark} &  & \\ \hline

\begin{tabular}[c]{@{}c@{}} Proposed \\ solution\end{tabular} & \multicolumn{1}{c|}{\cmark} & \multicolumn{1}{c|}{\cmark} &  \multicolumn{1}{c|}{\cmark} &\multicolumn{1}{c|}{} & \multicolumn{1}{c|}{} & \multicolumn{1}{c|}{\cmark} & \cmark & \multicolumn{1}{c|}{} & \multicolumn{1}{c|}{\cmark} & \cmark & \cmark \\ \hline
 
\end{tabular}%
}
\end{table*}

\subsection{Related Works}
Employing \gls{RIS} with the \gls{RSMA} and/or the power-domain \gls{NOMA} schemes offers an advantageous approach for enhancing \gls{PLS} in \gls{VLC} systems, enabling adaptive interference management and increased confidentiality. By strategically placing \gls{RIS} to manage light reflections, \gls{VLC} systems can expand coverage and minimize the risk of eavesdropping, thereby strengthening \gls{PLS} in various operational environments. The authors of~\cite{9756553} introduced an iterative Kuhn-Munkres algorithm to maximize the \gls{SSR} efficiently, demonstrating lower complexity and performance improvements. The results confirmed that \gls{RIS} can substantially enhance the security and performance of \gls{VLC} systems. A \gls{PLS} technique for a \gls{VLC} system enhanced by \gls{RIS} in the form of mirror array sheets was introduced in~\cite{9784887}. To maximize the \gls{SR}, beamforming weights and mirror orientations were optimized using a \gls{DRL} approach based on the \gls{DDPG} algorithm. The highest \gls{SR} was achieved, and an optimal mirror arrangement size was determined. The authors of~\cite{10511290} utilized \gls{RIS} to enhance \gls{PLS} through a hybrid system that integrated \gls{mmWave} and \gls{VLC}. Beamforming weights at \gls{VLC} and \gls{mmWave} access points, mirror array configurations, and phase shifts were optimized to increase the \gls{SR} while power constraints were considered. The \gls{DRL} approach was proposed using the \gls{DDPG} technique. The results demonstrated significant \gls{SR} enhancements, specifically in \gls{mmWave} scenarios, validating the approach's effectiveness. The \gls{PLS} of an indoor \gls{VLC} system was investigated in~\cite{10516681}, involving an intelligent reflecting mirror array. An average optical intensity constraint and average peak optical intensity constraints were explored, where a lower-bound \gls{SR} expression was derived. The authors of~\cite{9500409} investigated the use of intelligent mirror array-based \gls{RIS} to enhance the \gls{PLS} of \gls{VLC} systems. The objective was to maximize the \gls{SR} by finding the optimal mirror orientations using a particle swarm optimization algorithm. To enhance the \gls{PLS} in \gls{RIS}-aided \gls{VLC} systems, the time delay induced by the \gls{RIS} reflections was utilized in~\cite{10669590}. In addition, considering the \gls{RIS} assignment, an \gls{SR} maximization problem was formulated and solved using the \gls{GA}. In~\cite{10279487}, the \gls{SR} of the power-domain \gls{NOMA}-\gls{VLC} system was explored by utilizing \gls{RIS}. The findings revealed that the \gls{PLS} was improved via \gls{RIS} deployment. A Max-Min \gls{SR} problem for \gls{RIS}-aided \gls{NOMA}-based \gls{VLC} system was investigated in~\cite{10746540}. The problem considered optimizing the \glspl{LED} power allocation and the \gls{RIS}-to-user assignment. The problem was decoupled into two sub-problems, through the block coordinate descent algorithm, and the sub-problems were solved using semi-definite relaxation and successive convex approximation algorithms. Through simulations, the authors demonstrated the superiority of the proposed system in terms of the minimum \gls{SE} compared to counterpart systems with uniform power allocation, gain radio power allocation, uniform \gls{RIS} assignment, random \gls{RIS} assignment, and without \gls{RIS}.

\subsection{Contributions}
Table~\ref{Table_I} provides a comparison of the contributions of our work related to existing studies in terms of \gls{RIS}-aided secure \gls{VLC} systems. This comparison covers several key dimensions, including optimization decision variables, objective function, access technique, and receiver configuration. Existing \gls{RIS}-aided \gls{VLC} literature optimized either the orientation angles of the \gls{RIS} elements, or the association matrix that determines the \gls{LED}-\gls{RIS} element-user association relationship, or the VLC AP power allocation coefficients, or two of the three aforementioned decision variables. Alternatively, our work jointly optimizes these variables. This joint optimization framework achieves a precise evaluation of the \gls{RIS}-reflected \gls{VLC} optical channel to gain more control over the \gls{RIS} elements. Regarding the objective function, prior works have focused predominantly on maximization of the \gls{SR} and the \gls{SSR} while overlooking the minimum user rate requirement. In contrast, our work prioritizes the Max-Min of both the \gls{SR} and the \gls{SEE} to ensure fair, robust, and secure communication. Regarding access techniques, while previous research has applied \gls{OMA} schemes or the power-domain \gls{NOMA} scheme, our approach considers both the \gls{RSMA} and the power-domain \gls{NOMA} schemes, enhancing spectral efficiency and user fairness by accommodating various network requirements. Additionally, incorporating random device orientation in the receiver configuration extends our model's practical applicability and robustness. Finally, the proposed approach jointly optimizes power allocation, \gls{RIS} association, and \gls{RIS} elements orientation angles. These advancements significantly elevate the \gls{PLS} of \gls{RSMA} and power-domain \gls{NOMA}-based \gls{VLC} systems, marking a substantial contribution to \gls{RIS}-enabled \gls{PLS} in \gls{VLC} systems. The primary contributions of this work are summarized as follows:

\begin{itemize}
\item We propose two optimization frameworks that maximize both the minimum \gls{SR} and the minimum \gls{SEE} by jointly optimizing power allocation, \gls{RIS} association, and \gls{RIS} elements orientation angles for both power-domain \gls{NOMA} and \gls{RSMA}-based \gls{VLC} systems. The two optimization frameworks ensure a fair, robust, and secure \gls{VLC} environment. In addition, the proposed joint optimization framework achieves high-precision control of \gls{RIS} elements, increasing flexibility and the degrees of freedom in the design of \gls{RIS}-aided \gls{VLC} systems.

\item We employ the \gls{RSMA} and the power-domain \gls{NOMA} schemes for more significant flexibility in managing network resources, improving spectral efficiency by enabling simultaneous access for multiple users, and enhancing user fairness by adjusting resource allocation to ensure a minimum rate requirement of each user. 

\item We extend the practicality and robustness of our model by incorporating the random device orientation in the receiver configuration. Thus, practical deployment scenarios can be evaluated by accommodating variations in device orientation, which can significantly impact system performance. 

\item Comprehensive simulation results are presented to show that the \gls{RSMA} scheme outperforms the power-domain \gls{NOMA} scheme in the proposed system, based on both of the considered \gls{SR} and \gls{SEE} metrics. This comparison takes into account various network parameters, including the \gls{VLC} \gls{AP} electrical transmit power, the \gls{FoV} of the \gls{PD}, the number of deployed \gls{RIS} elements, the reflectivity of the \glspl{RIS}, the minimum rate requirements of each user, and the number of served users. Additionally, we present simulation results related to the convergence and benchmarking of the proposed solution methodology to illustrate its efficacy. 

\end{itemize}

The remainder of this paper is structured as follows: Section~\ref{Sec: System and Channel Models} details the system and channel models of both legitimate users and the eavesdropper. Section~\ref{Sec: Adopted Multiple Access Schemes} introduces the multiple access schemes employed in this study. The optimization problems are formulated in Section~\ref{Sec: Problem Formulation}, followed by the solution methodology in Section~\ref{Sec: Solution Methodology}. Section~\ref{Sec: Simulation Results} discusses the simulation results. Finally, the paper's conclusion is summarized in Section~\ref{Sec: Conclusion}.

\begin{figure*}[!t]
\centering
\includegraphics[width=\textwidth]{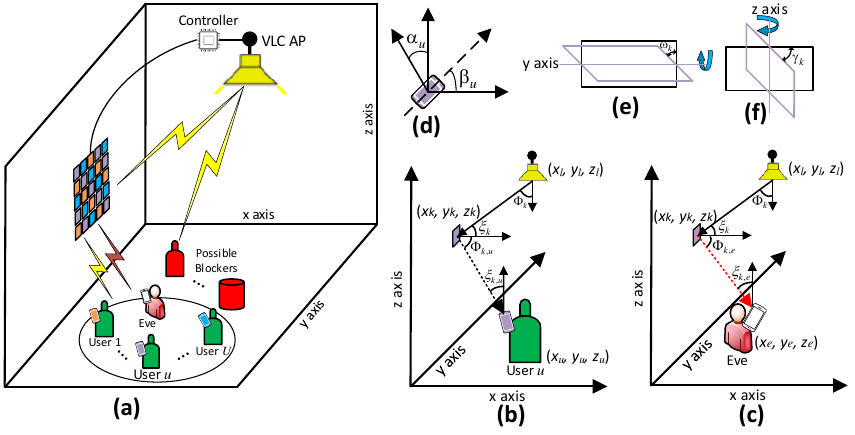}
\caption{(a) An illustration of the proposed mirror array-based \gls{RIS}-aided secure \gls{VLC} system, (b) the angles of irradiance and incidence of the light signal that propagates from the \gls{VLC} \gls{AP} through the $k$-th mirror array-based \gls{RIS} element to the $u$-th user, (c) the eavesdropped light signal while specifying the angles of incidence and irradiance of this signal, (d) the user device's orientation in respect of the device's polar and azimuth angles ($\alpha_u,\beta_u$), (e) the mirror array-based \gls{RIS} orientation in respect of its roll angle $\omega_k$, and (f) the mirror array-based \gls{RIS} orientation in respect of its yaw angle $\gamma_k$.
}
\label{fig: System_Model}
\end{figure*}

\section{System and Channel Models}
\label{Sec: System and Channel Models}

\subsection{Indoor VLC System}
The system model of the considered mirror array-based \gls{RIS}-aided secure \gls{VLC} system is illustrated in Fig.~\ref{fig: System_Model}(a), with a focus on downlink communications. In this figure, a \gls{VLC} \gls{AP} mounted on the ceiling serves multiple legitimate users via mirror array-based \glspl{RIS} positioned on the room's wall. Due to the presence of objects and/or unintended users (i.e., possible blockers) in the room, we consider a \gls{NLoS} path between the \gls{VLC} \gls{AP} and the intended users. In addition to the legitimate users, there is an eavesdropper, Eve, who attempts to wiretap the secret message for the legitimate users from the reflected signals of the mirror array-based \glspl{RIS}. It is desired to optimize the assignment of the \glspl{RIS} elements to the legitimate users, the orientation of the mirror array-based \glspl{RIS}, and the transmit power of the \gls{VLC} \gls{AP} such that the \gls{PLS} of the system is maximized. Through channel acquisition methods, the controller unit can obtain the spatial locations (and subsequently, the required channel information) of legitimate users, Eve, \gls{RIS} elements, and the \gls{VLC} \gls{AP}~\cite{10746540, 7747504, 9784887, 10279487}. Fig.~\ref{fig: System_Model}(b) and Fig.~\ref{fig: System_Model}(c) show the angles of irradiance and incidence of the light signal that propagates from the \gls{VLC} \gls{AP} through the $k$-th mirror array-based \gls{RIS} element to both the $u$-th user and Eve, respectively. For which, $u \in [1,\dots,U]$ and $k \in [1,\dots,K]$ denote an arbitrary user and an arbitrary \gls{RIS} element, respectively. $U$ and $K$ denote the number of intended users and the number of \gls{RIS} elements, respectively.  Fig.~\ref{fig: System_Model}(d) illustrates the user's device orientation angles, while Fig.~\ref{fig: System_Model}(e) and Fig.~\ref{fig: System_Model}(f) show the mirror array-based \gls{RIS} orientation in respect of its roll angle $\omega_k$ and yaw angle $\gamma_k$, respectively. The channel gains for the legitimate users and Eve are elaborated in the sequel.

\subsection{VLC Channel of Legitimate Users}

The \gls{NLoS} channel gain between the \gls{AP} and the $u$-th user, considering the signal reflection from the $k$-th \gls{RIS} element can be expressed as~\cite{9276478}
\begin{equation} \label{eq: The channel gain of the NLoS link - RIS to the user - diffuse}
    h^{\textnormal{NLoS}}_{k,u}= \begin{cases} \rho_{\textnormal{RIS}}\frac{(m+1)A_{\textnormal{PD}}}{2\pi^2 d_{k}^2 d_{k,u}^2} {A}_k G_c(\xi) G_f(\xi) \cos^m(\Phi_{k}) \times \\ \cos(\xi_{k}) \cos(\Phi_{k,u}) \cos(\xi_{k,u}), \ 0 \leq \xi_{k,u} \leq \xi_{\textnormal{FoV}} \\
    0, \ \xi_{k,u} > \xi_{\textnormal{FoV}} \end{cases}
\end{equation}
where $\rho_{\textnormal{RIS}}$ is the reflectivity of the \glspl{RIS}, $m=-\log_2 \left( \cos \left(\Phi_{1/2} \right) \right)$ denotes the order of Lambertian emission with the semi-angle at half power of LEDs $\Phi_{1/2}$, $A_{\textnormal{PD}}$ represents the physical area of the \gls{PD}, ${A}_k$ is the area of the $k$-th reflective surface, $G_f(\xi)$ is the gain of the optical filter, $\Phi_{k}$ is the irradiance angle between the \gls{AP} and the $k$-th \gls{RIS} element, $\xi_{k}$ is the incidence angle between the \gls{AP} and the $k$-th \gls{RIS} element, $\Phi_{k,u}$ is the irradiance angle between the $k$-th \gls{RIS} element and the $u$-th user, $\xi_{k,u}$ is the incidence angle between the $k$-th \gls{RIS} element and the $u$-th user, and $G_c(\xi)$ is the gain of the optical concentrator which is given by $G_c(\xi)=f^2/ \sin^2{\xi}_{\textnormal{FoV}}$, where $f$ and ${\xi}_{\textnormal{FoV}}$ are the refractive index and \gls{FoV} of the \gls{PD}, respectively. The variables $d_{k}^2$ and $d_{k,u}^2$ denote the link distance between the \gls{VLC} \gls{AP} and the $k$-th \gls{RIS} element and the $k$-th \gls{RIS} element and the $u$-th user, respectively. 

The orientation of the user's device does not affect the irradiance angles $\Phi_{k}$ and $\Phi_{k,u}$. However, the incidence angle $\xi_{k,u}$ is heavily impacted by device orientation. In (\ref{eq: The channel gain of the NLoS link - RIS to the user - diffuse}), the term $\cos(\xi_{k,u})$ captures the impact of random device orientation on the signal reception at the \gls{PD} and can be expressed as~\cite{8540452}
\begin{equation} \label{eq: Cos Xi}
\begin{split}
\cos(\xi_{k,u}) = & \big(\frac{x_k - x_u}{d_{k,u}}\big) \cos(\beta_u) \sin(\alpha_u) + \\ & \big(\frac{y_k - y_u}{d_{k,u}}\big) \sin(\beta_u) \sin(\alpha_u) + \\ & \big(\frac{z_k - z_u}{d_{k,u}}\big) \cos(\alpha_u), 
\end{split}
\end{equation}
where $\left(x_k,y_k,z_k\right)$ and $\left(x_u, y_u, z_u\right)$ are the Cartesian coordinates for the $k$-th \gls{RIS} element and the  $u$-th user, respectively. $\alpha_u$ and $\beta_u$ are the polar and azimuth angles of the $u$-th user device, respectively. Considering the orientation angles of the $k$-th \gls{RIS} element that has been assigned to the $u$-th user (i.e., $\gamma_k$ and $\omega_k$), the cosine of the irradiance angle $\Phi_{k,u}$ can be expressed as~\cite{9543660}   

\begin{equation} \label{eq: Cos Phi}
\begin{split}
\cos(\Phi_{k,u})= & \big(\frac{x_k - x_u}{d_{k,u}}\big) \sin(\gamma_k) \cos(\omega_k) + \\
& \big(\frac{y_k - y_u}{d_{k,u}}\big) \cos(\gamma_k) \cos(\omega_k) + \\
& \big(\frac{z_k - z_u}{d_{k,u}}\big) \sin(\omega_k).
\end{split}
\end{equation}

\subsection{VLC Channel of Eve}

We assume that Eve has a powerful receiver and can eavesdrop on all the users in the system~\cite{9464264,9270035}. The channel gain of the eavesdropped signal of the $u$-th user can be expressed as
\begin{equation} \label{eq: The channel gain of the NLoS link - RIS to Eve - diffuse}
    h^{\textnormal{NLoS}}_{k,e}= \begin{cases} \rho_{\textnormal{RIS}}\frac{(m+1)A_{\textnormal{PD}}}{2\pi^2 d_{k}^2 d_{k,e}^2} {A}_k G_c(\xi) G_f(\xi) \cos^m(\Phi_{k}) \times \\ \cos(\xi_{k}) \cos(\Phi_{k,e}) \cos(\xi_{k,e}), \ 0 \leq \xi_{k,e} \leq \xi_{\textnormal{FoV}} \\
    0, \ \xi_{k,e} > \xi_{\textnormal{FoV}} \end{cases}
\end{equation}
where $\cos(\Phi_{k,e})$ and $\cos(\xi_{k,e})$ are given by 

\begin{equation} \label{eq: Cos Phi of Eve}
\begin{split}
\cos(\Phi_{k,e})= & \big(\frac{x_k - x_e}{d_{k,e}}\big) \sin(\gamma_k) \cos(\omega_k) + \\
& \big(\frac{y_k - y_e}{d_{k,e}}\big) \cos(\gamma_k) \cos(\omega_k) + \\
& \big(\frac{z_k - z_e}{d_{k,e}}\big) \sin(\omega_k), 
\end{split}
\end{equation}
and

\begin{equation} \label{eq: Cos Xi of Eve}
\begin{split}
\cos(\xi_{k,e}) = & \big(\frac{x_k - x_e}{d_{k,e}}\big) \cos(\beta_e) \sin(\alpha_e) + \\ & \big(\frac{y_k - y_e}{d_{k,e}}\big) \sin(\beta_e) \sin(\alpha_e) + \\ & \big(\frac{z_k - z_e}{d_{k,e}}\big) \cos(\alpha_e), 
\end{split}
\end{equation}
respectively.

\section{Adopted Multiple Access Schemes}
\label{Sec: Adopted Multiple Access Schemes}

In this section, the details of the considered $1$-layer \gls{RSMA} scheme and the power-domain \gls{NOMA} scheme are presented. In the considered multi-user scenario, the users are sorted based on their channel gains 
$h^{\textnormal{NLoS}}_{k,1} \geq \dots \geq h^{\textnormal{NLoS}}_{k,U}$~\cite{maraqa2021achievable}. To facilitate the analysis, a matrix and a vector are defined, respectively, as follows 
\begin{equation} \label{RIS-user association matrix}
    \textbf{G} = [\textbf{g}_1, \textbf{g}_2,\dots, \textbf{g}_u]_{K \times U},
\end{equation}

\begin{equation} \label{eq: Channel vector from all associated RIS elements}
    \textbf{h}^{\textnormal{NLoS}}_{u} = [h^{\textnormal{NLoS}}_{1,u}, h^{\textnormal{NLoS}}_{2,u},\dots, h^{\textnormal{NLoS}}_{K,u}],
\end{equation}
\noindent where $\textbf{G}$ denotes the \gls{RIS}-user association matrix and $\textbf{g}_u$ is a vector that indicates the \gls{RIS} elements assigned to the $u$-th user. For example, $g_{k,u} \in \{0,1\}$  is a component of $\textbf{g}_u$, representing an indicator function that determines whether or not the $k$-th \gls{RIS} element is assigned to the $u$-th user. In this work, we assume that Eve is an internal Eve (i.e., untrusted user), and thus can know the signaling structure of the system to be able to eavesdrop on the $u$-th user signal~\cite{9032167,9780419}.

\subsection{RSMA Scheme}

The \gls{RSMA} scheme divides each user's message into two parts: a private part and a common part. At the \gls{VLC} \gls{AP} side, all the common parts are combined into a single message, which is then encoded into a common stream. This common stream is broadcasted to all users along with each user's private stream. On the user side, all users can decode the common stream, but only the intended user can decode its own private stream. More specifically, using the \gls{SIC} technique, users can remove the common stream from their received signal, allowing them to decode their private stream while treating other users' private streams as noise~\cite{9831440}. At the $u$-th user, the received signal can be expressed as
\begin{equation} \label{eq: received signal RSMA}
    y_{u}^\textnormal{RSMA} = \big(\textbf{h}^{\textnormal{NLoS}}_{u}\big)^T \textbf{g}_u  \times (\sqrt{P_\textnormal{0}} s_c + \sum_{u=1}^{U} \sqrt{P_u} s_u) + z_u,
\end{equation} 
\noindent where $P_\textnormal{0}$ and $P_u$ are the transmit power coefficients of the common stream and the private stream, respectively. Those transmit power coefficients are continuous variables and the relation between them is given by $P_\textnormal{S} = P_\textnormal{0} + \sum_{u=1}^{U} P_u$, where $P_\textnormal{S}$ is the \gls{VLC} \gls{AP} electrical transmit power, while $s_c$ and $s_u$ denote the $u$-th user common and private streams, respectively. $z_u \sim \mathcal{N}(0,\sigma^2)$ is the real-valued additive Gaussian noise with variance $\sigma^2$ and it includes both the thermal and the shot noises~\cite{10375270}.

At the $u$-th user, the message decoding process starts with decoding the common stream and subtracting it from the message through the \gls{SIC} process. Accordingly, the rate of the $u$-th user decoding the common stream is given as \cite{10375270}
\begin{equation} \label{eq: rate of user u decoding common stream}
\begin{split}
R_{c,u}^\textnormal{RSMA} = & B \textnormal{log}_2 \Bigg(1 + \frac{\textnormal{exp(1)}}{2 \pi} \frac{\big( R_{\textnormal{PD}} \big(\textbf{h}^{\textnormal{NLoS}}_{i}\big)^T \textbf{g}_i \big)^2 P_\textnormal{0}}{ I^\textnormal{Common}_u + N_o B}  \Bigg), \\ & 1 \leq i \leq U 
\end{split}
\end{equation}
\noindent where $I^\textnormal{Common}_u=( R_{\textnormal{PD}} \big(\textbf{h}^{\textnormal{NLoS}}_{i}\big)^T \textbf{g}_i)^2 \sum_{j=1}^U P_j$ represents the inter-user interference term resulting from implementing the \gls{RSMA} scheme. $R_{\textnormal{PD}}$, $N_o$, and $B$ denote the \gls{PD} responsivity, the noise power spectral density, and the bandwidth of the system, respectively. 

At each user, after subtracting the common stream from the message, the user decodes the private stream while treating other users' private streams as noise. The rate of decoding the private stream is
\begin{equation} \label{eq: achievable rate of user u decoding private stream}
\begin{split}
R_{p,u}^\textnormal{RSMA} = & B \textnormal{log}_2 \Bigg(1 + \frac{\textnormal{exp(1)}}{2 \pi} \frac{\big( R_{\textnormal{PD}} \big(\textbf{h}^{\textnormal{NLoS}}_{i}\big)^T \textbf{g}_i \big)^2 P_i}{ I_u^\textnormal{Private} + N_o B}  \Bigg), \\ & 1 \leq i \leq U 
\end{split}
\end{equation}
\noindent where $I_u^\textnormal{Private}=( R_{\textnormal{PD}} \big(\textbf{h}^{\textnormal{NLoS}}_{i}\big)^T \textbf{g}_i )^2 \sum_{j=1, \ j \neq u}^U P_j$ represents the inter-user interference term resulting from implementing the \gls{RSMA} scheme.

Eve operates similarly to a legitimate user by first decoding the common stream, subtracting it from the eavesdropped user's message, and subsequently attempting to decode the private user message. Accordingly, the achievable rate of decoding the common stream of the $u$-th user at Eve is

\begin{equation} \label{eq: rate of eve decoding common stream}
\begin{split}
R_{c,e\rightarrow u}^\textnormal{RSMA} = & B \textnormal{log}_2 \Bigg(1 + \frac{\textnormal{exp(1)}}{2 \pi} \frac{\big( R_{\textnormal{PD}} \big(\textbf{h}^{\textnormal{NLoS}}_{e}\big)^T \textbf{g}_i \big)^2 P_\textnormal{0}}{ I^\textnormal{Common}_e + N_o B}  \Bigg), \\ & 1 \leq i \leq U
\end{split}
\end{equation}
\noindent where $I^\textnormal{Common}_e=( R_{\textnormal{PD}} \big(\textbf{h}^{\textnormal{NLoS}}_{e}\big)^T \textbf{g}_i)^2 \sum_{j=1}^U P_j$ represents the inter-user interference term resulting from implementing the \gls{RSMA} scheme. likewise, the achievable rate of decoding the private stream of the $u$-th user at Eve is given by
\begin{equation} \label{eq: achievable rate of Eve decoding private stream}
\begin{split}
R_{p,e\rightarrow u}^\textnormal{RSMA} = & B \textnormal{log}_2 \Bigg(1 + \frac{\textnormal{exp(1)}}{2 \pi} \frac{\big( R_{\textnormal{PD}} \big(\textbf{h}^{\textnormal{NLoS}}_{e}\big)^T \textbf{g}_i \big)^2 P_i}{ I^\textnormal{Private}_e + N_o B}  \Bigg), \\ & 1 \leq i \leq U 
\end{split}
\end{equation}
\noindent where $I^\textnormal{Private}_e=( R_{\textnormal{PD}} \big(\textbf{h}^{\textnormal{NLoS}}_{e}\big)^T \textbf{g}_i )^2 \sum_{j=1, \ j \neq u}^U P_j$ represents the inter-user interference term resulting from implementing the \gls{RSMA} scheme. In general, the \gls{SR} can be determined by calculating the difference between the achievable rates of the legitimate user and the eavesdropper.  Mathematically, the \gls{SR} for multiple users can be defined as~\cite{wang2016physical} 
\begin{equation}    
\textnormal{SR}^{\upsilon}=\left [ {R}_{u}-{R}_{e \to u} \right]^{+},
\end{equation}
where $ \upsilon\in \left\{ \textrm{RSMA}, \textrm{NOMA}\right\} $, ${R}_{u}$ is the achievable rate of the legitimate user,  ${R}_{e\to u}$ is the achievable rate of the eavesdropper, wiretapping the $u$-th user, and $\left [x\right ]^{+}=\max\left ( x,0 \right )$. In this work, we first aim to maximize the minimum secrecy rate, which for the proposed \gls{RSMA} \gls{VLC} system is given by~\cite{9195771}
\begin{equation} \label{eq: min secrecy RSMA}
  \textnormal{SR}_\textnormal{min}^{\textnormal{RSMA}} = \min_{u = [1,\ldots,U]}[(R_{c,u}^{\textnormal{RSMA}}+R_{p,u}^{\textnormal{RSMA}})-(R_{c,e\rightarrow u}^{\textnormal{RSMA}}+R_{p,e\rightarrow u}^{\textnormal{RSMA}})]^+. 
\end{equation}

For clarity,~\eqref{eq: min secrecy RSMA} is expanded at the bottom of this page.

\subsection{Power-domain NOMA Scheme}
\label{subSec: Power-domain NOMA Scheme}
The power-domain \gls{NOMA} scheme allows several users to share the same radio resource. At the \gls{VLC} \gls{AP} side, through the principle of superposition coding, a superposed signal is formed and transmitted to the intended users. On the user side, through the principle of \gls{SIC}, users' messages can be decoded~\cite{9154358}. Subsequently, at the $u$-th user, the received signal can be expressed as
\begin{equation} \label{eq: Received signal u-th user}
    y_{u}^{\textnormal{NOMA}} =  \big(\textbf{h}^{\textnormal{NLoS}}_{u}\big)^T \textbf{g}_u \times (\sum_{u=1}^{U} \sqrt{c_u P_\textnormal{S}} s_u) + z_u,
\end{equation} 
\noindent where $c_u$ denotes the \gls{NOMA} power allocation coefficient of the $u$-th user. Based on~\cite{shen2023secrecy}, $c_u$ is given as
\begin{equation} \label{eq: allocated power ratio}
c_u =  \begin{cases}
       \epsilon^{\textnormal{NOMA}} (1-\epsilon^{\textnormal{NOMA}})^{u-1}, \ \textnormal{if} \ 1 \leq u<U \\
       (1-\epsilon^{\textnormal{NOMA}})^{u-1}, \ \textnormal{if} \ u=U
       \end{cases}
\end{equation}
\noindent where $\epsilon^{\textnormal{NOMA}}$ is a constant value in the range of $(0.5,1]$. Also, $\epsilon^{\textnormal{NOMA}}$ satisfies the following relation $\sum_{u=1}^U c_u~=~1$. 

Accordingly, through the power-domain \gls{NOMA} scheme, the decoding rate at the $u$-th user is given as
\begin{equation} \label{eq: u-th user rate}
\begin{aligned}
& R_{u}^{\textnormal{NOMA}} = \begin{cases} B \textnormal{log}_2 \Bigg(1 + \frac{\textnormal{exp(1)}}{2 \pi} \frac{\big( R_{\textnormal{PD}} \big(\textbf{h}^{\textnormal{NLoS}}_{u}\big)^T \textbf{g}_u \big)^2 c_u P_\textnormal{S}}{N_o B}  \Bigg), u=1, \\
B \textnormal{log}_2 \Bigg(1 + \frac{\textnormal{exp(1)}}{2 \pi}  \frac{\big( R_{\textnormal{PD}} \big(\textbf{h}^{\textnormal{NLoS}}_{u}\big)^T \textbf{g}_u  \big)^2 c_u P_\textnormal{S}}{ I_u + N_o B}  \Bigg), \\ 1 < u \leq U,
\end{cases}
\end{aligned}
\end{equation}
\noindent where $I_u= \big( R_{\textnormal{PD}} \big(\textbf{h}^{\textnormal{NLoS}}_{u}\big)^T \textbf{g}_u \big)^2 \sum_{i=1}^{u-1} c_i P_\textnormal{S}$ represents the inter-user interference term resulting from implementing the power-domain \gls{NOMA} scheme.

The Eve's data rate when the $u$-th user is being wiretapped can be expressed as 
\begin{equation} \label{eq: EVE rate}
\begin{aligned}
& R_{e\rightarrow u}^{\textnormal{NOMA}} = \begin{cases} B \textnormal{log}_2 \Bigg(1 + \frac{\textnormal{exp(1)}}{2 \pi} \frac{\big( R_{\textnormal{PD}} \big(\textbf{h}^{\textnormal{NLoS}}_{e}\big)^T \textbf{g}_u \big)^2 c_u P_\textnormal{S}}{N_o B}  \Bigg), u=1,\\
B \textnormal{log}_2 \Bigg(1 + \frac{\textnormal{exp(1)}}{2 \pi}  \frac{\big( R_{\textnormal{PD}} \big(\textbf{h}^{\textnormal{NLoS}}_{e}\big)^T \textbf{g}_u  \big)^2 c_u P_\textnormal{S}}{I_e + N_o B}  \Bigg), \\ 1 < u \leq U,
\end{cases}
\end{aligned}
\end{equation}
\noindent where $\textbf{h}^{\textnormal{NLoS}}_{e} = [h^{\textnormal{NLoS}}_{1,e}, h^{\textnormal{NLoS}}_{2,e},\dots, h^{\textnormal{NLoS}}_{K,e}]$ and $I_e= \big( R_{\textnormal{PD}} \big(\textbf{h}^{\textnormal{NLoS}}_{e}\big)^T \textbf{g}_u \big)^2 \sum_{i=1}^{u-1} c_i P_\textnormal{S}$. Thus, the minimum secrecy rate of the proposed \gls{NOMA} \gls{VLC} system can be expressed as~\cite{8713985}
\begin{equation} \label{eq: min secrecy NOMA}
  \textnormal{SR}_\textnormal{min}^{\textnormal{NOMA}} = \min_{u = [1,\ldots,U]} [R_{u}^{\textnormal{NOMA}}-R_{e\rightarrow u}^{\textnormal{NOMA}}]^+,
\end{equation}
\noindent where for clarity,~\eqref{eq: min secrecy NOMA} is expanded at the top of the next page.

\begin{strip}
\begin{bottomborder} \end{bottomborder}

\begin{flalign*}
\textnormal{SR}_\textnormal{min}^{\textnormal{RSMA}} &= \min_{u = [1,\ldots,U]}[(R_{c,u}^{\textnormal{RSMA}}+R_{p,u}^{\textnormal{RSMA}})-(R_{c,e\rightarrow u}^{\textnormal{RSMA}}+R_{p,e\rightarrow u}^{\textnormal{RSMA}})]^+                                                                                                                    && \nonumber \\ & = \min_{u = [1,\ldots,U]} \Bigg[ \Big(B \textnormal{log}_2 \Bigg(1 + \frac{\textnormal{exp(1)}}{2 \pi}  \frac{\big( R_{\textnormal{PD}} \big(\textbf{h}^{\textnormal{NLoS}}_{u}\big)^T \textbf{g}_u \big)^2 P_\textnormal{0}}{\big( R_{\textnormal{PD}} \big(\textbf{h}^{\textnormal{NLoS}}_{u}\big)^T \textbf{g}_u \big)^2 \sum_{j=1}^U P_j + N_o B}  \Bigg)+                                    
&& \nonumber \\ & \qquad \qquad \quad \ \ B \textnormal{log}_2 \Bigg(1 + \frac{\textnormal{exp(1)}}{2 \pi} \frac{\big( R_{\textnormal{PD}} \big(\textbf{h}^{\textnormal{NLoS}}_{u}\big)^T \textbf{g}_u \big)^2 P_u}{ ( R_{\textnormal{PD}} \big(\textbf{h}^{\textnormal{NLoS}}_{u}\big)^T \textbf{g}_u )^2 \sum_{j=1, \ j \neq u}^U P_j + N_o B}  \Bigg) \Big)-                          
&& \nonumber \\ & \qquad \qquad \quad \ \Big(B \textnormal{log}_2 \Bigg(1 + \frac{\textnormal{exp(1)}}{2 \pi} \frac{\big( R_{\textnormal{PD}} \big(\textbf{h}^{\textnormal{NLoS}}_{e}\big)^T \textbf{g}_u \big)^2 P_\textnormal{0}}{ ( R_{\textnormal{PD}} \big(\textbf{h}^{\textnormal{NLoS}}_{e}\big)^T \textbf{g}_u)^2 \sum_{j=1}^U P_j + N_o B}  \Bigg)+                                     && \nonumber \\ & \qquad \qquad \quad \ \  B \textnormal{log}_2 \Bigg(1 + \frac{\textnormal{exp(1)}}{2 \pi} \frac{\big( R_{\textnormal{PD}} \big(\textbf{h}^{\textnormal{NLoS}}_{e}\big)^T \textbf{g}_u \big)^2 P_u}{ ( R_{\textnormal{PD}} \big(\textbf{h}^{\textnormal{NLoS}}_{e}\big)^T \textbf{g}_u )^2 \sum_{j=1, \ j \neq u}^U P_j + N_o B}  \Bigg) \Big) \Bigg]^+,
\end{flalign*}

\begin{flalign*}
\textnormal{SR}_\textnormal{min}^{\textnormal{NOMA}} &= \min_{u = [1,\ldots,U]} [R_{u}^{\textnormal{NOMA}}-R_{e\rightarrow u}^{\textnormal{NOMA}}]^+ && \nonumber \\ & = \min_{u = [1,\ldots,U]} \Bigg[B \textnormal{log}_2 \Bigg(1 + \frac{\textnormal{exp(1)}}{2 \pi}  \frac{\big( R_{\textnormal{PD}} \big(\textbf{h}^{\textnormal{NLoS}}_{u}\big)^T \textbf{g}_u  \big)^2 c_u P_\textnormal{S}}{ \big( R_{\textnormal{PD}} \big(\textbf{h}^{\textnormal{NLoS}}_{u}\big)^T \textbf{g}_u \big)^2 \sum_{i=1}^{u-1} c_i P_\textnormal{S} + N_o B}  \Bigg)- && \nonumber \\ & \qquad \qquad \quad \ B \textnormal{log}_2 \Bigg(1 + \frac{\textnormal{exp(1)}}{2 \pi}  \frac{\big( R_{\textnormal{PD}} \big(\textbf{h}^{\textnormal{NLoS}}_{e}\big)^T \textbf{g}_u  \big)^2 c_u P_\textnormal{S}}{ \big( R_{\textnormal{PD}} \big(\textbf{h}^{\textnormal{NLoS}}_{e}\big)^T \textbf{g}_u \big)^2 \sum_{i=1}^{u-1} c_i P_\textnormal{S} + N_o B}  \Bigg)\Bigg]^+.
\end{flalign*}

\begin{bottomborder} \end{bottomborder}
\end{strip}

\section{Problem Formulation}
\label{Sec: Problem Formulation}
In this section, the details of the Max-Min secrecy rate optimization problem of both the \gls{RSMA} and the power-domain \gls{NOMA} schemes for the proposed combined channel model are presented. This optimization problem simultaneously optimizes several optimizations variables, (i) the \gls{VLC} \gls{AP} power allocation coefficients (this variable is represented in the \gls{RSMA} scheme by $P_0$ and $P_u$, and in the power-domain \gls{NOMA} scheme by $c_u$), (ii) the \gls{RIS} elements orientation angles (this variable is represented for both considered schemes by the angles $\omega_k$ and $\gamma_k$), and (iii) the \gls{RIS} association matrix (this variable is represented for both considered schemes by the matrix $\textbf{G}$).

\subsection{Max-Min Secrecy Rate Optimization}
One can express the Max-Min secrecy rate optimization problem of the \gls{RSMA} scheme as
\begin{alignat}{3}
\textnormal{(P1):} &\max_{\{\textbf{G}, \omega_k,\gamma_k, P_0, P_u \}} & & \textnormal{SR}_\textnormal{min}^{\textnormal{RSMA}},\label{eq:objective_RSMA}\\
& \qquad \ \text{s.t.} &  &   g_{k,u} \in \{0,1\}, \ \forall k \in [1,..., K],  \nonumber \\ &  &  & \forall u \in [1,..., U],   \label{eq:c1_RSMA}\\
& &  &  \sum_{u=1}^{U} g_{k,u}=1, \ \forall k\in [1,..., K], \label{eq:c2_RSMA} \\
& &  & -\frac{\pi}{2} \leq \omega_k \leq \frac{\pi}{2}, \label{eq:c3_RSMA} \\
& &  & -\frac{\pi}{2} \leq \gamma_k \leq \frac{\pi}{2}, \label{eq:c4_RSMA}\\ 
& &  &  P_0 + \sum_{u=1}^{U} P_u \leq P_S \label{eq:c5_RSMA},\\
& &  &  R_{c,u}^{\textnormal{RSMA}} + R_{p,u}^{\textnormal{RSMA}} \geq \bar{R}_{\textnormal{min}}, \ \forall u \in [1,..., U], \label{eq:c6_RSMA}
\end{alignat}
\noindent where the constraint~\eqref{eq:c1_RSMA} comes from the definition of each entity, $g_{k,u}$, in the \gls{RIS}-user association matrix, $\textbf{G}$, which was previously described in~\eqref{RIS-user association matrix}. The constraint~\eqref{eq:c2_RSMA} ensures that each \gls{RIS} element can only serve a particular user at a time. This constraint follows the point source assumption proposed in~\cite{9276478} and is utilized in the context of \gls{RIS}-aided \gls{VLC} systems in several works, for example,~\cite{9756553}. The constraints~\eqref{eq:c1_RSMA} and~\eqref{eq:c2_RSMA} control the \gls{RIS}-user association decision variable, the first decision variable in (P1), by setting a binary indicator function to each \gls{RIS} element to serve a particular user at a time. The constraints~\eqref{eq:c3_RSMA} and~\eqref{eq:c4_RSMA} ensure the  orientation angles, $\omega_k$ and $\gamma_k$, lie within the range of $[-\frac{\pi}{2},\frac{\pi}{2}]$. The goal of these two constraints is to adjust the orientation of each \gls{RIS} element to steer the light toward the intended receiver. The constraint~\eqref{eq:c5_RSMA} denotes the \gls{VLC} \gls{AP} total electrical transmit power limitation. This constraint is set to ensure that the \gls{VLC} \gls{AP} adheres to this power limitation, in addition to efficiently distributing this power between the common and private streams in the adopted \gls{RSMA} scheme. The constraint~\eqref{eq:c6_RSMA} guarantees the minimum rate requirements of each user, $\bar{R}_{\textnormal{min}}$. Consequently, this constraint ensures a certain \gls{QoS} for all users in the system.

Similarly, one can express the Max-Min secrecy rate optimization problem of the power-domain \gls{NOMA} scheme as
\begin{alignat}{3}
\textnormal{(P2):} &\max_{\{\textbf{G}, \omega_k,\gamma_k, c_u \}} &\quad& \textnormal{SR}_\textnormal{min}^{\textnormal{NOMA}},\label{eq:objective_NOMA}\\
&\qquad \ \text{s.t.} &  &   g_{k,u} \in \{0,1\}, \ \forall k \in [1,..., K], \nonumber \\ &  &  & \forall u \in [1,..., U],   \label{eq:c1_NOMA}\\
& &  &  \sum_{u=1}^{U} g_{k,u}=1, \ \forall k\in [1,..., K], \label{eq:c2_NOMA} \\
& &  &  -\frac{\pi}{2} \leq \omega_k \leq \frac{\pi}{2}, \label{eq:c3_NOMA} \\
& &  &  -\frac{\pi}{2} \leq \gamma_k \leq \frac{\pi}{2}, \label{eq:c4_NOMA}\\ 
& &  &  \sum_{u=1}^{U} c_u \leq 1 \label{eq:c5_NOMA}, \\
& &  &  R_{u}^{\textnormal{NOMA}} \geq \bar{R}_{\textnormal{min}}, \ \forall u \in [1,..., U], \label{eq:c6_NOMA}
\end{alignat}
\noindent where the constraints~\eqref{eq:c5_NOMA} make sure that the allocated power ratios to each user in the power-domain \gls{NOMA} scheme are within the \gls{VLC} \gls{AP} total electrical transmit power limitation. The constraint~\eqref{eq:c6_NOMA} guarantees the minimum rate requirements of each user, $\bar{R}_{\textnormal{min}}$.

Both (P1) and (P2) are novel optimization problems that have not been considered in the literature yet. These problems are non-convex and include both discrete (i.e., $\textbf{G}$) and continuous (i.e., $\omega_k$, $\gamma_k$, $P_0$, $P_u$ and $c_u$) optimization variables, classifying them as \gls{MINLP} NP-hard problems. Also, the number of optimization variables that need to be optimized in each of these optimization problems is high. The complexity of (P1) and (P2) are presented in Section V-B. This raises the problems to a greater level of complexity, making them difficult to be solved using conventional methods. The typical approach for such joint optimization problems is by decomposing it into sub-problems. However, due to the structure of the considered objective functions and the multiple mixed integer decision variables, decomposing (P1) and (P2) still yields non-convex sub-problems. Furthermore, such a decomposition approach affects the quality of the solution since the decision variables are not jointly optimized. Therefore, to solve (P1) and (P2) within a reasonable time while ensuring solution quality, we resort to evolutionary metaheuristic algorithms, specifically the \gls{GA}. It is worth noting that the \gls{GA} is popular in solving NP-hard problems. This is because this algorithm combines global search capabilities, flexibility, and robustness, which are critical when tackling NP-hard problems with vast and complex solution spaces~\cite{Sait:1999}. Also, the \gls{GA} often provides an acceptable solution (i.e., a solution that respects the constraints) within a reasonable time, making it highly practical for complex optimization problems. Moreover, the \gls{GA} has shown efficiency in solving \gls{RIS}-related optimization problems, as highlighted in~\cite{10445725, 10500850, 10669590}. The details of the solution methodology are presented in Section~\ref{Sec: Solution Methodology}. 

\subsection{Max-Min Secrecy Energy Efficiency Optimization}
Alongside the secrecy rate, the \gls{SEE} is a popular performance metric in \gls{PLS}-aided \gls{VLC} systems~\cite{9463422, 9633190, 10404069}. The \gls{SEE} metric ensures that the consumed energy of the system is taken into consideration while optimizing the secrecy performance of the system. The \gls{SEE} can be defined as the ``consumed energy per transmitted confidential bits (Bits/Joule)''~\cite{10404069}. Generally, there are three main contributors to the total consumed energy in \gls{RIS}-aided \gls{VLC} system, namely, the \gls{VLC} \gls{AP} (i.e., Transmitter), \gls{RIS}, and users (i.e., Receivers)~\cite{8307185}. Subsequently, the total consumed energy in the proposed system is $P_{\textnormal{Total}} = P_\textnormal{Transmitter} + P_\textnormal{RIS} + P_\textnormal{Receviers} = (P_\textnormal{S} + P_\textnormal{DAC} + P_\textnormal{Filter} + P_\textnormal{PA} + P_\textnormal{Driver} + P_\textnormal{T-Circuit}) + (P_\textnormal{Element} \times K) + U*(P_\textnormal{ADC} + P_\textnormal{TIA} + P_\textnormal{Filter} + P_\textnormal{R-Circuit})$. At the transmitter, $P_\textnormal{DAC}$, $P_\textnormal{Filter}$, $P_\textnormal{PA}$, $P_\textnormal{Driver}$, and $P_\textnormal{T-Circuit}$ denote the consumed energy related to the digital-to-analog converter (DAC), the filter, the power amplifier, the \gls{LED} driver, and the transmitter external circuit, respectively. At the \gls{RIS}, $P_\textnormal{Element}$ denotes the consumed energy that is needed to rotate each passive \gls{RIS} element. At each user, $P_\textnormal{ADC}$, $P_\textnormal{TIA}$, $P_\textnormal{Filter}$, and $P_\textnormal{R-Circuit}$ denote the consumed energy related to the analog-to-digital converter (ADC), the trans-impedance amplifier (TIA), the filter, and the receiver external circuit, respectively. The minimum secrecy rate of the proposed \gls{RIS}-aided \gls{VLC} system for both the \gls{RSMA} and the power-domain \gls{NOMA} scheme are provided in~\eqref{eq: min secrecy RSMA} and~\eqref{eq: min secrecy NOMA}, respectively. Therefore, the minimum \gls{SEE} for both the \gls{RSMA} and the power-domain \gls{NOMA} scheme can be expressed respectively as      
\begin{equation} \label{eq: SEE RSMA}
    \textnormal{SEE}_\textnormal{min}^\textnormal{RSMA} = \frac{\textnormal{SR}_\textnormal{min}^{\textnormal{RSMA}}}{P_{\textnormal{Total}}},
\end{equation}

\begin{equation} \label{eq: SEE NOMA}
      \textnormal{SEE}_\textnormal{min}^\textnormal{NOMA} = \frac{\textnormal{SR}_\textnormal{min}^{\textnormal{NOMA}}}{P_{\textnormal{Total}}}.
\end{equation}
\indent To formulate the Max-Min \gls{SEE} problem for the considered \gls{RSMA} scheme, we replace the objective functions of (P1) by~\eqref{eq: SEE RSMA}. Similarly, for the considered \gls{NOMA} scheme, we replace the objective functions of (P2) by~\eqref{eq: SEE NOMA}. The Max-Min \gls{SEE} optimization problems can be expressed as

\begin{alignat}{3}
\textnormal{(P3):} &\max_{\{\textbf{G}, \omega_k,\gamma_k, P_0, P_u \}} & & \textnormal{SEE}_\textnormal{min}^\textnormal{RSMA},\label{eq:objectiveSEE_RSMA}\\
&  &  &  \eqref{eq:c1_RSMA}, \ \eqref{eq:c2_RSMA}, \ \eqref{eq:c3_RSMA}, \ \eqref{eq:c4_RSMA}, \ \eqref{eq:c5_RSMA}, \ \eqref{eq:c6_RSMA},  \nonumber 
\end{alignat}
\noindent and 
\begin{alignat}{3}
\textnormal{(P4):} &\max_{\{\textbf{G}, \omega_k,\gamma_k, c_u \}} & & \textnormal{SEE}_\textnormal{min}^\textnormal{NOMA},\label{eq:NOMA}\\
&  &  &  \eqref{eq:c1_NOMA}, \ \eqref{eq:c2_NOMA}, \ \eqref{eq:c3_NOMA}, \ \eqref{eq:c4_NOMA}, \ \eqref{eq:c5_NOMA}, \ \eqref{eq:c6_NOMA}.  \nonumber 
\end{alignat}

Then, we solve (P3) and (P4) by following the same steps that will be mentioned in Section~\ref{Sec: Solution Methodology}. It is worth noting that (P3) and (P4) have fractional objective functions, which adds another level of complexity compared to the involved objective functions in (P1) and (P2). This is an additional reason for considering the \gls{GA} as a unified solution methodology for all these optimization problems.

\section{Solution Methodology}
\label{Sec: Solution Methodology}

\subsection{Genetic Algorithm-based Approach}

The working principle behind the \gls{GA} is the idea of the ``survival of the fittest", whereby the best individuals in an initial randomly generated solution candidates (population of chromosomes) are selected and combined to form the next generation of chromosomes. In our case, an individual can be a \gls{RIS}-user association variable, \gls{RIS} elements orientation angles, and power allocation coefficients for the \gls{RSMA} and the power-domain \gls{NOMA} schemes. After a predefined number of generations or if the maximum fitness value is reached, the evolution process is stopped, and the surviving chromosomes represent the best solution~\cite{mitchell1998}. Particularly, three types of operators are involved, namely, selection, crossover, and mutation. The selection operator selects a pair of chromosomes according to the tournament method for selecting the best individuals. The crossover operator randomly chooses a locus and exchanges the sub-sequences before and after that locus between the two chromosomes to create two offspring. The mutation operator flips some of the bits in a chromosome to help prevent the algorithm from getting stuck in a local optima. After generating the two offspring, their fitness and feasibility indicators are evaluated. The two offspring replace some chromosomes in the current population according to the following adopted replacement procedure. The two chromosomes with the lowest fitness value and zero feasibility indicator are replaced. If all the chromosomes in the population are feasible, the two chromosomes with the lowest fitness values are replaced.     

The \gls{GA}-based algorithm for a population size ${\mathcal J}$, number of generations $N_{\textnormal{Gen}}$, and probabilities of crossover and mutation of ${\mathcal P}_c$ and ${\mathcal P}_m$, respectively, is summarized in Algorithm~{\ref{Algorithm: proposed solution 1}}.

\SetAlgoNlRelativeSize{-1}
\begin{algorithm}[!t]
\scriptsize
\DontPrintSemicolon
\SetAlgoLined
\caption{Proposed GA-based solution.} \label{Algorithm: proposed solution 1}
\LinesNumbered
\KwIn{Population size ${\mathcal J}$, number of generations ${N_{\textnormal{Gen}}}$, ${\mathbf{h}}^{\textnormal{NLoS}}_u$, ${\mathbf{h}}^{\textnormal{NLoS}}_e$, $\bar{R}_{\textnormal{min}}$, $P_S$;}
\KwOut{Optimized solutions $\textbf{G}, \omega_k, \gamma_k, c_u, P_0, P_u$;}
Generate an initial population of $\mathcal{J}$ randomly constructed chromosomes ${\mathcal Y}_0$;\;
\For{$n=1:{N_{\textnormal{Gen}}}$}{
Select a pair of parent chromosomes from ${\mathcal Y}_{n-1}$;\;
With crossover probability ${\mathcal P}_c$, apply crossover operation on selected pair to form two new offspring;\;
With mutation probability ${\mathcal P}_m$, apply mutation operation on the two offspring to form a new feasible chromosome;\;
Place the resulting chromosome in the updated population ${\mathcal Y}_{n}$;\; 
Evaluate the fitness of each chromosome in ${\mathcal Y}_{n}$ using the appropriate objective function;\;
Select elite chromosomes;\;
}
Update ${\mathcal Y}_{\textnormal{best}}$;\;
\end{algorithm}

\begin{table}[!t]
\centering
\caption{Simulation Parameters}
\label{tab: Simulation Parameters}
\resizebox{0.5\textwidth}{!}{%
\begin{tabular}{|l|l|c|}
\hline
Parameter name, notation & Value & [\#] \\ \hline

The VLC AP electrical transmit power, $P_S$ & $5$ Watt & \cite{8669970} \\

The number of RIS elements, $K$ & $100$ & \cite{10375270}\\ 

The FoV of the PD, $\xi_{\textnormal{FoV}}$ & $85^{\circ}$ & \cite{maraqa2021achievable}\\ 

The reflection coefficient of the RIS element, $\rho_\textnormal{RIS}$  & $0.95$ & \cite{9543660}\\

The half-intensity radiation angle, $\Phi_{1/2}$ & $70^\circ$& \cite{10183987}\\

The physical area of the PD, $G_f(\xi)$ & $1.0$ & \cite{10959088}\\

The refractive index of the PD, $f$ & $1.5$& \cite{10375270}\\

The responsivity of the PD, $R_\textnormal{PD}$ & $0.53$ A/W& \cite{10183987}\\  

The bandwidth of the system, $B$ & $200$ MHz & \cite{10959088}\\

The minimum rate requirements of each user, $\bar{R}_{\textnormal{min}}$  & $30$ kbps& \cite{rasti2021design}\\

The population size in the Genetic algorithm, $\mathcal{J}$ & $150$& \cite{Sait:1999}\\

The number of generations in the Genetic algorithm, $N_\textnormal{Gen}$ & $100$ & \cite{Sait:1999}\\ 

The noise power spectral density, $N_o$ & $10^{-21} \textnormal{A}^2/\textnormal{Hz}$ & \cite{maraqa2021achievable}\\ 

The number of intended users, $U$ & $4$ & \cite{10375270}\\

The power of the filter, $P_\textnormal{Filter}$ & $2.5$ mWatt & \cite{10183987}\\ 

The power of the DAC, $P_\textnormal{DAC}$ & $175$ mWatt & \cite{10183987}\\ 

The power of the LED driver, $P_\textnormal{Driver}$ & $2758$ mWatt & \cite{10183987}\\ 

The power of the power amplifier, $P_\textnormal{PA}$ & $280$  mWatt & \cite{10183987} \\  

The power of the transmitter external circuit, $P_\textnormal{T-Circuit}$ & $3250$ mWatt & \cite{10183987} \\

The consumed energy for rotating an RIS element, $P_\textnormal{Element}$ & $100$  mWatt & \cite{10183987} \\  

The consumed energy of the ADC, $P_\textnormal{ADC}$ & $95$ mWatt & \cite{10183987}\\

The consumed energy of the TIA, $P_\textnormal{TIA}$ & $2500$ mWatt & \cite{10183987} \\  

The consumed energy of the receiver external circuit, $P_\textnormal{R-Circuit}$ & $1.9$ mWatt & \cite{10183987} \\ 

\hline
\end{tabular}
}
\end{table}

\subsection{Computational Complexity}
\label{subsection: Computational Complexity}

The computational complexity of Algorithm~\ref{Algorithm: proposed solution 1} is composed of the following components:

\begin{itemize}
    \item Initialization: The time complexity for generating an ${\mathcal S}$ random chromosomes or individuals is ${\mathcal O}\left(  {\mathcal S}\right)$, where ${\mathcal S}$ is the number of decision variables. In both considered schemes, to optimize $\textbf{G}$, the required number of decision variables is $U*K$ and to optimize $\omega_k$ and $\gamma_k$, the required number of decision variables is $2*K$. In the \gls{RSMA} scheme, to optimize $P_0$ and $P_u$, the required number of decision variables is $1+U$. On the other hand, to optimize $c_u$ in the power-domain \gls{NOMA} scheme, the required number of decision variables is one. Overall, $\mathcal{S}^\textnormal{RSMA}=(U*K)+(2*K)+(1+U)$ and $\mathcal{S}^\textnormal{NOMA}=(U*K)+(2*K)+(1)$ for \gls{RSMA} and power-domain \gls{NOMA} schemes, respectively. For example, if $K$ is $100$ and $U$ is $4$, then the required number of decision variables are $605$ and $601$ for the \gls{RSMA} and the power-domain \gls{NOMA} schemes, respectively. It is worth noting that in our proposed system, the power-domain \gls{NOMA} scheme requires fewer decision variables compared to the \gls{RSMA} scheme. Consequently, it has less computational complexity.
    \item Selection: The time complexity of the tournament selection method is ${\mathcal O}\left(  2{\mathcal S}\right)$ \cite{de2019batch}. 
    \item Crossover and Mutation: The crossover and mutation processes each has a worst-case time complexity of ${\mathcal S}*O_l$, where $O_l$ is the locus (i.e., crossover point).
    \item Evaluation: The evaluation of the fitness function for all individuals can be done in parallel and hence has a time complexity ${\mathcal O}(1)$.
\end{itemize}
Accordingly, the overall time complexity can be approximated as ${\mathcal O}\left(S\right) + {\mathcal O}\left(2S\right) + {\mathcal O}\left(S*O_l\right) \approx {\mathcal O}\left(S*O_l\right)$.

\section{Simulation Results}
\label{Sec: Simulation Results}

This section provides detailed numerical results to assess the secrecy performance of the proposed \gls{RIS}-aided \gls{VLC} system for the considered \gls{RSMA} and power-domain \gls{NOMA} schemes in terms of both the Max-Min \gls{SR} and the \gls{SEE}. The default parameters for the simulations are summarized in Table~\ref{tab: Simulation Parameters}. Beyond this summary, the polar angle, $\alpha_u$, is characterized using the Laplace distribution and its value spans over $[0,\frac{\pi}{2}]$ with a mean of $41^\circ$ and a standard deviation of $9^\circ$~\cite{8540452}. The azimuth angle, $\beta_u$, is characterized using the uniform distribution, and its value spans over $[-\pi,\pi]$~\cite{8540452}. The users in the system are depicted as cylinders with a height of $1.65$ meters and a radius of $0.15$ meters. The receiver is held by a user standing at a height of $0.85$ meters and a distance of $0.36$ meters from their body. The default size of the considered \gls{RIS} is $10 \ \textnormal{rows} \times 10 \ \textnormal{columns}$, while each element in the \gls{RIS} is a square with dimensions of $10$ cm $\times$ $10$ cm. The dimensions of the room are assumed to be $5$ m $\times$ $5$ m $\times$ $3$ m. 

\begin{figure}[!b]
\centering
\vspace{-2em}
\includegraphics[width=0.485\textwidth]{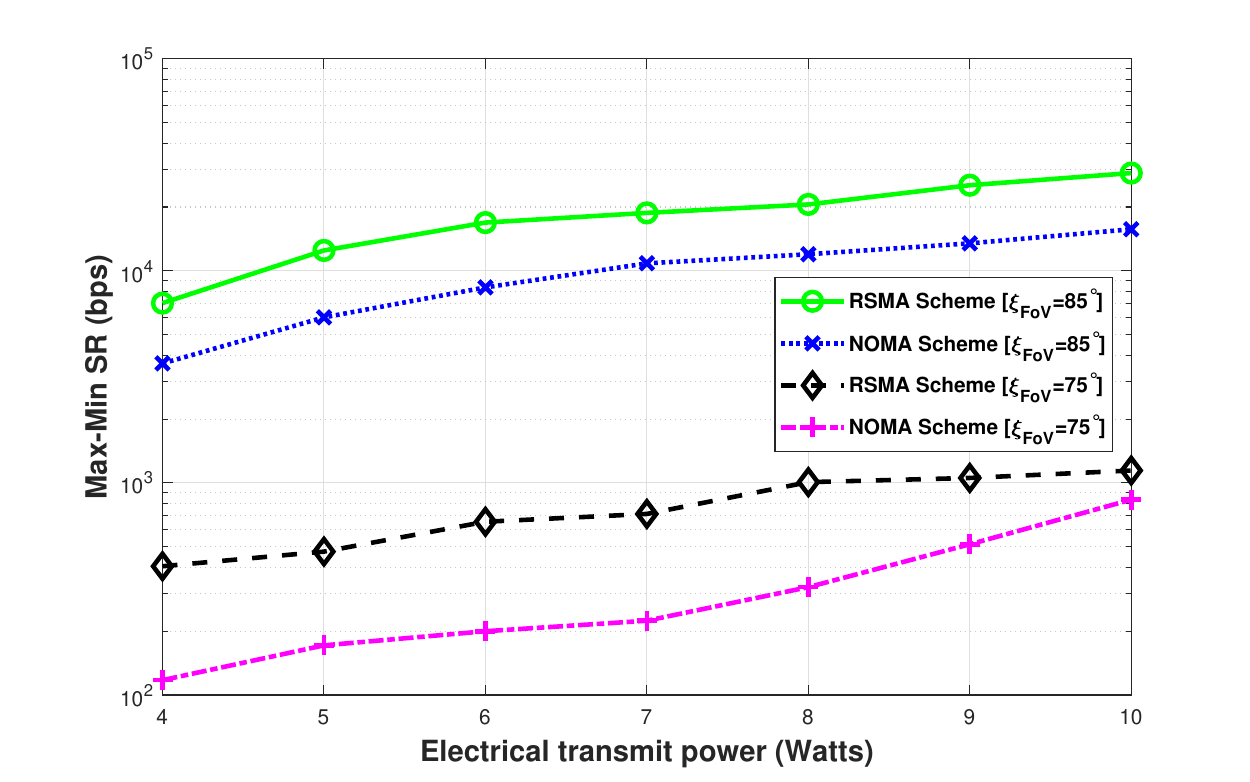}
\caption{Max-Min \gls{SR} performance versus \gls{VLC} \gls{AP} electrical transmit power for different \glspl{PD} \gls{FoV}.}
\label{fig: A_Tx_Power_vs_Rate_new}
\end{figure}

\subsection{Max-Min Secrecy Rate Results}

Fig.~\ref{fig: A_Tx_Power_vs_Rate_new} shows that the proposed \gls{RIS}-aided \gls{VLC} system with the \gls{RSMA} scheme achieves a better Max-Min \gls{SR} performance compared to the power-domain \gls{NOMA} scheme. This result indicates that the \gls{RSMA} scheme is more secure for \gls{VLC} users than the power-domain \gls{NOMA} scheme. Specifically, under the same network parameters, a better \gls{SR} is achieved by the \gls{RSMA} scheme. Also, this figure shows the effect of changing the \glspl{PD} \gls{FoV} on the Max-Min \gls{SR} performance. It is known that the wider a user \gls{PD} \gls{FoV}, the better the chance that the user receives stronger light signals from the closest \gls{RIS} elements. Consequently, this figure illustrates that even a $10^{\circ}$ increase in the users \gls{PD} \gls{FoV} can achieve up to $42$-fold enhancement in the \gls{SR} performance.

\begin{figure}[!t]
\centering
\vspace{-2em}
\includegraphics[width=0.485\textwidth]{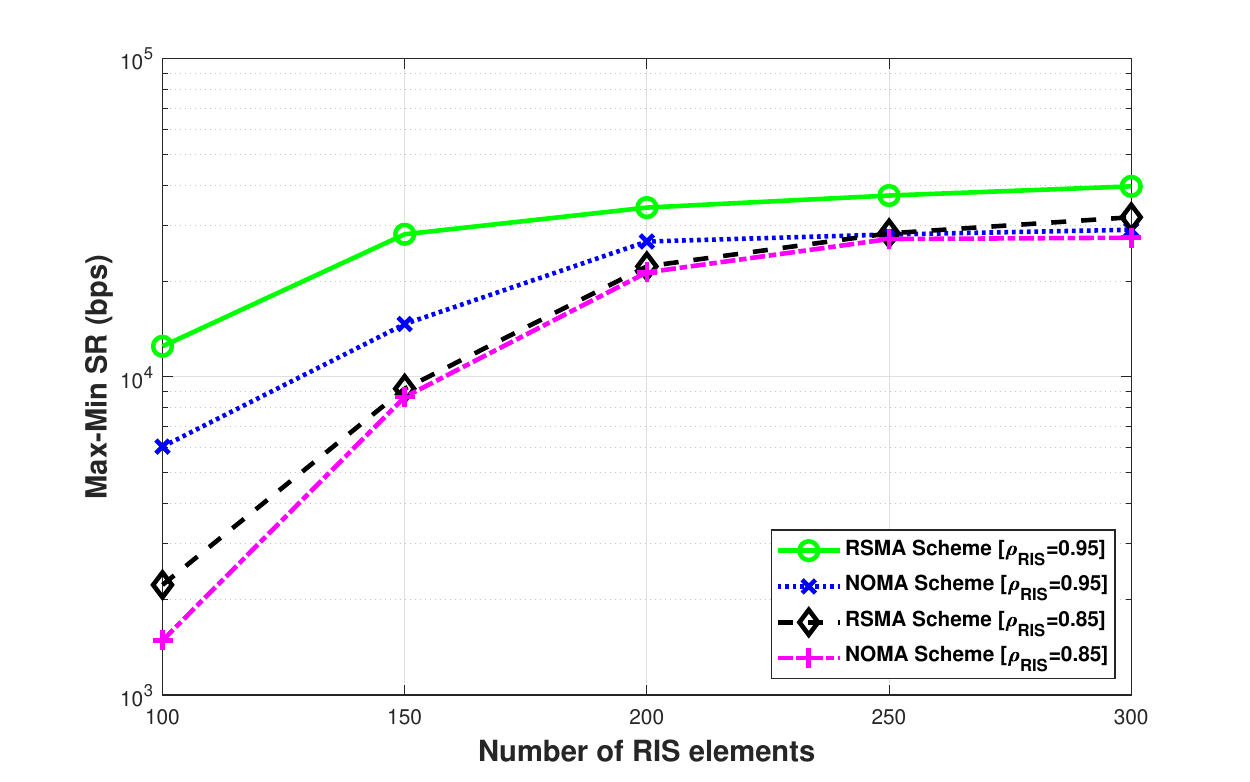}
\caption{Max-Min \gls{SR} performance versus number of \gls{RIS} elements for different \glspl{RIS} reflectivities.}
\label{fig: B_Elements_vs_SR_new}
\end{figure}

\begin{figure}[!t]
\centering
\vspace{-2em}
\includegraphics[width=0.485\textwidth]{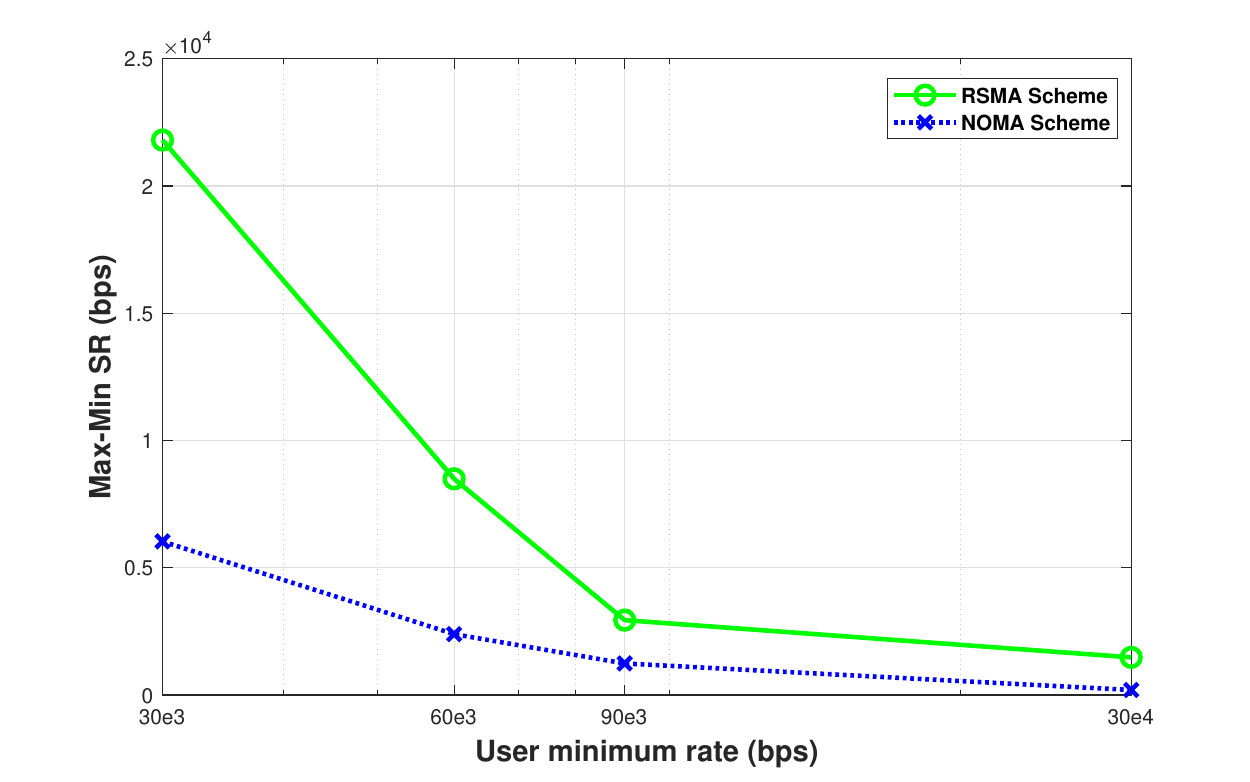}
\caption{Max-Min \gls{SR} performance versus users' minimum rate requirement.}
\label{fig: C_Rmin_vs_SR}
\end{figure}

\begin{figure}[!t]
\centering
\vspace{-1.5em}
\includegraphics[width=0.485\textwidth]{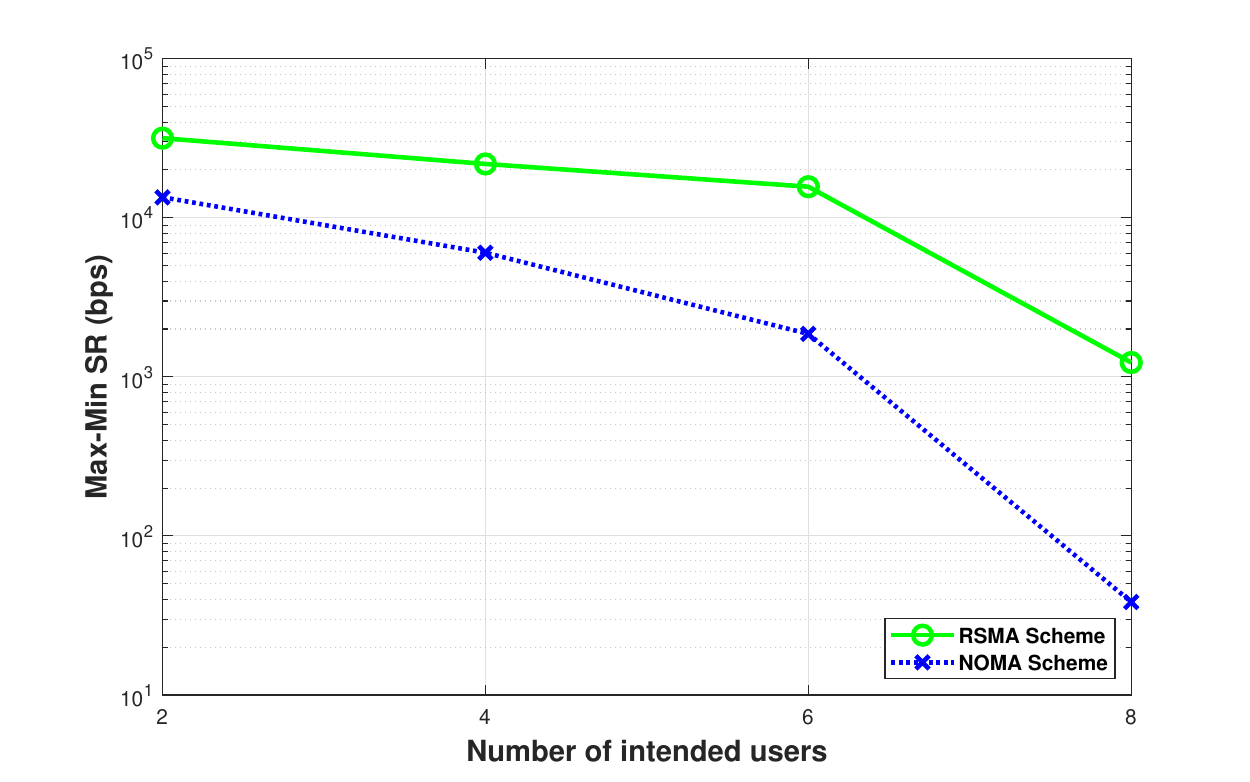}
\caption{Max-Min \gls{SR} performance versus number of intended users.}
\label{fig: D_Vusers}
\end{figure}

Fig.~\ref{fig: B_Elements_vs_SR_new} compares the Max-Min \gls{SR} performance of the proposed \gls{RIS}-aided \gls{VLC} system for both the \gls{RSMA} and the power-domain \gls{NOMA} schemes. As expected, as the number of \gls{RIS} elements increases, the Max-Min \gls{SR} performance for both schemes increases. It can be observed that this gain increases steadily and then starts to saturate with a high number of elements. This is supported by the fact that adding more elements to a large \gls{RIS} leads to only marginal improvements. Also, this figure shows the effect of changing the \glspl{RIS} reflectivities, $\rho_{\textnormal{RIS}}$. For best performance, it is important to choose \gls{RIS} elements with high reflectivities, as a $10\%$ decrease in the elements' reflectivity can degrade the \gls{SR} performance by at least $18\%$.

Fig.~\ref{fig: C_Rmin_vs_SR} demonstrates the Max-Min \gls{SR} performance versus users' minimum rate requirement. In this figure, as the required rate $\bar{R}_\textnormal{min}$ increases, more resources are used to satisfy the per-user demand, leading to a decrease in the Max-Min \gls{SR}. This observation is true for both the \gls{RSMA} and the power-domain \gls{NOMA} schemes.

\begin{figure}[!t]
\centering
\vspace{-2em}
\includegraphics[width=0.485\textwidth]{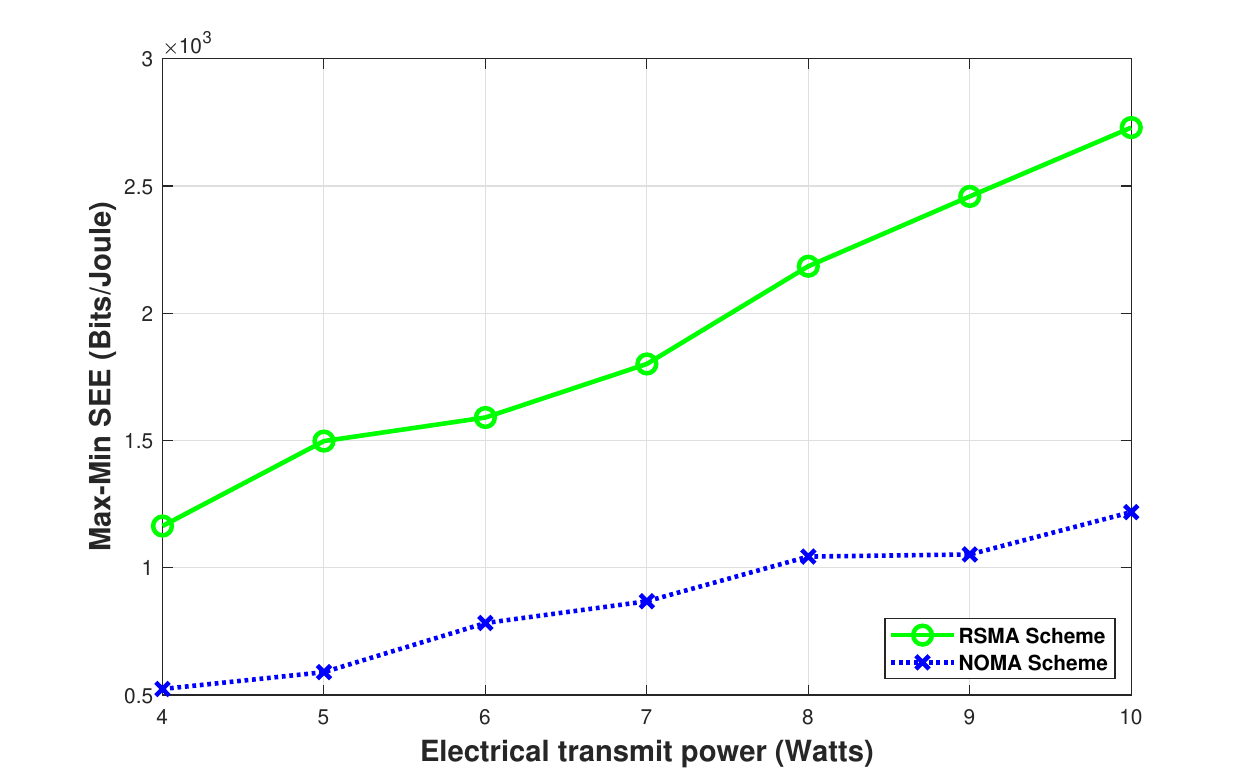}
\caption{Max-Min \gls{SEE} performance versus \gls{VLC} \gls{AP} electrical transmit power.}
\label{fig: E_Tx_Power_vs_SEE}
\end{figure}

Fig.~\ref{fig: D_Vusers} illustrates the Max-Min \gls{SR} performance versus the number of intended users. It can be observed that as the number of users in the system increases, the Max-Min \gls{SR} performance decreases. This can be justified by the fact that the number of resources in the considered \gls{RIS}-aided \gls{VLC} system is fixed (i.e., the number of \gls{RIS} elements that provide intended users with light signals is fixed). 

\subsection{Max-Min Secrecy Energy Efficiency Results} 

Fig.~\ref{fig: E_Tx_Power_vs_SEE} compares the Max-Min \gls{SEE} performance of the proposed \gls{RIS}-aided \gls{VLC} system for the \gls{RSMA} and the power-domain \gls{NOMA} schemes for different values of the \gls{VLC} \gls{AP} electrical transmit power. One can notice that the \gls{RSMA} scheme achieves a significant performance gain, up to $254\%$, compared to the power-domain \gls{NOMA} scheme. This is because the \gls{RSMA} scheme is more energy efficient compared to the power-domain \gls{NOMA} scheme~\cite{8491100}.   

\begin{figure}[!t]
\centering
\vspace{-1.5em}
\includegraphics[width=0.485\textwidth]{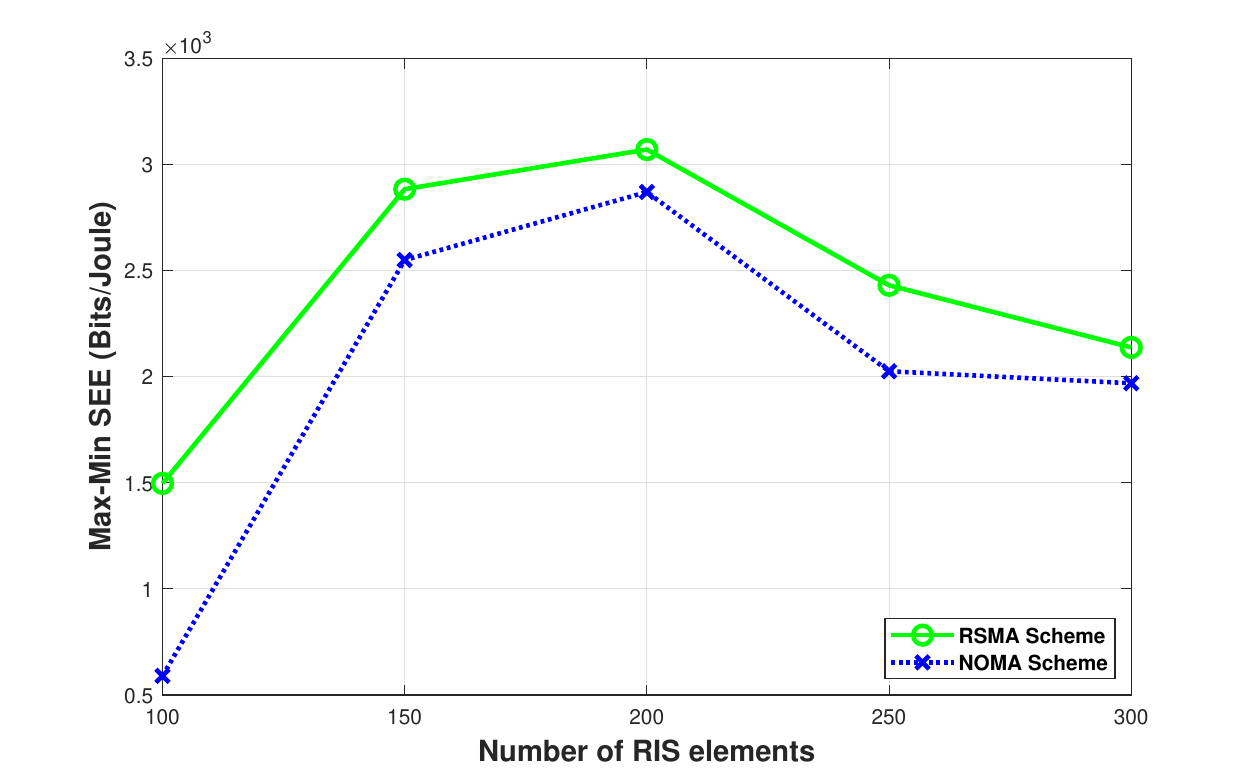}
\caption{Max-Min \gls{SR} performance versus number of \gls{RIS} elements.}
\label{fig: F_elements_vs_SEE}
\end{figure}

\begin{table*}[!t]
\centering
\vspace{-1em}
\caption{A compact example of the optimized decision variables for the evaluated Max-Min \gls{SR} optimization of the proposed \gls{RSMA}-based \gls{RIS}-aided \gls{VLC} system.}
\vspace{-0.5em}
\label{tab: optimized parameters}
\resizebox{\textwidth}{!}{%
\begin{tabular}{|c|ccc||c|cccc|}
\hline

\multirow{3}{*}{RIS \#} & \multicolumn{2}{c|}{\begin{tabular}[c]{@{}c@{}} RIS Association $\textbf{G}$ \end{tabular}} & \multicolumn{1}{c|}{\multirow{2}{*}{\begin{tabular}[c]{@{}c@{}} RIS Orientation \\ ($\omega_{1:15},\gamma_{1:15}$) \end{tabular}}} & \multirow{3}{*}{RIS \#} & \multicolumn{2}{c|}{\begin{tabular}[c]{@{}c@{}} RIS Association $\textbf{G}$ \\ (Cont.) \end{tabular}} & \multicolumn{1}{c|}{\multirow{2}{*}{\begin{tabular}[c]{@{}c@{}} RIS Orientation \\ ($\omega_{16:30},\gamma_{16:30}$) \\ (Cont.) \end{tabular}}} & \multirow{2}{*}{\begin{tabular}[c]{@{}c@{}} VLC AP \\ power allocation \\ ($P_0, P_1, P_2$) \end{tabular}} \\[4mm] \cline{2-3} \cline{6-7}
 
& \multicolumn{1}{c|}{User 1} & \multicolumn{1}{c|}{User 2} & \multicolumn{1}{c|}{} & & \multicolumn{1}{c|}{User 1} & \multicolumn{1}{c|}{User 2} & \multicolumn{1}{c|}{} &  \\ \hline

1 & \multicolumn{1}{c|}{0} & \multicolumn{1}{c|}{1} & \multicolumn{1}{c|}{(23.0891\textdegree, 15.5374\textdegree)} & 16 & \multicolumn{1}{c|}{0} & \multicolumn{1}{c|}{1} & \multicolumn{1}{c|}{(76.5314\textdegree, 72.8200\textdegree)} & \multirow{15}{*}{\begin{tabular}[c]{@{}c@{}} (0.7152, 0.1,\\ 0.1796)\end{tabular}} \\ \cline{1-8}

2 & \multicolumn{1}{c|}{0} & \multicolumn{1}{c|}{1} & \multicolumn{1}{c|}{(52.5489\textdegree, -18.6762\textdegree)} & 17 & \multicolumn{1}{c|}{1} & \multicolumn{1}{c|}{0} & \multicolumn{1}{c|}{(40.6991\textdegree, 58.2423\textdegree)} & \multicolumn{1}{c|}{} \\ \cline{1-8}

3 & \multicolumn{1}{c|}{0} & \multicolumn{1}{c|}{1} & \multicolumn{1}{c|}{(41.8877\textdegree, 52.6141\textdegree)} & 18 & \multicolumn{1}{c|}{0} & \multicolumn{1}{c|}{1} & \multicolumn{1}{c|}{(42.4476\textdegree, -74.5671\textdegree)} & \multicolumn{1}{c|}{} \\ \cline{1-8}

4 & \multicolumn{1}{c|}{0} & \multicolumn{1}{c|}{1} & \multicolumn{1}{c|}{(36.4535\textdegree, -1.3691\textdegree)} & 19 & \multicolumn{1}{c|}{1} & \multicolumn{1}{c|}{0} & \multicolumn{1}{c|}{(-35.7331\textdegree, -5.9635\textdegree)} & \multicolumn{1}{c|}{} \\ \cline{1-8}

5 & \multicolumn{1}{c|}{0} & \multicolumn{1}{c|}{1} & \multicolumn{1}{c|}{(18.9253\textdegree, -48.5019\textdegree)} & 20 & \multicolumn{1}{c|}{1} & \multicolumn{1}{c|}{0} & \multicolumn{1}{c|}{(33.2268\textdegree, 55.9919\textdegree)} & \multicolumn{1}{c|}{} \\ \cline{1-8}

6 & \multicolumn{1}{c|}{0} & \multicolumn{1}{c|}{1} & \multicolumn{1}{c|}{(-22.4593\textdegree, 43.0449\textdegree)} & 21 & \multicolumn{1}{c|}{0} & \multicolumn{1}{c|}{1} & \multicolumn{1}{c|}{(-16.5819\textdegree, 69.9737\textdegree)} & \multicolumn{1}{c|}{} \\ \cline{1-8}

7 & \multicolumn{1}{c|}{0} & \multicolumn{1}{c|}{1} & \multicolumn{1}{c|}{(-31.6412\textdegree, -5.4666\textdegree)} & 22 & \multicolumn{1}{c|}{0} & \multicolumn{1}{c|}{1} & \multicolumn{1}{c|}{(-18.4175\textdegree, -24.0222\textdegree)} & \multicolumn{1}{c|}{} \\ \cline{1-8}

8 & \multicolumn{1}{c|}{0} & \multicolumn{1}{c|}{1} & \multicolumn{1}{c|}{(-33.4945\textdegree, -50.6330\textdegree)} & 23 & \multicolumn{1}{c|}{0} & \multicolumn{1}{c|}{1} & \multicolumn{1}{c|}{(71.5116\textdegree, -32.2981\textdegree)} &  \multicolumn{1}{c|}{} \\ \cline{1-8}

9 & \multicolumn{1}{c|}{0} & \multicolumn{1}{c|}{1} & \multicolumn{1}{c|}{(78.7889\textdegree, 61.0648\textdegree)} & 24 &  \multicolumn{1}{c|}{0} & \multicolumn{1}{c|}{1} & \multicolumn{1}{c|}{(69.8503\textdegree, 8.6299\textdegree)} & \multicolumn{1}{c|}{} \\ \cline{1-8}

10 & \multicolumn{1}{c|}{0} & \multicolumn{1}{c|}{1} & \multicolumn{1}{c|}{(-33.1080\textdegree, -29.6117\textdegree)} & 25 & \multicolumn{1}{c|}{1} & \multicolumn{1}{c|}{0} & \multicolumn{1}{c|}{(-80.1327\textdegree, -78.3159\textdegree)} & \multicolumn{1}{c|}{} \\ \cline{1-8}

11 & \multicolumn{1}{c|}{0} & \multicolumn{1}{c|}{1} & \multicolumn{1}{c|}{(-74.2463\textdegree, 75.4771\textdegree)} & 26 & \multicolumn{1}{c|}{0} & \multicolumn{1}{c|}{1} & \multicolumn{1}{c|}{(-34.0116\textdegree, -56.7357\textdegree)} & \multicolumn{1}{c|}{} \\ \cline{1-8}

12 & \multicolumn{1}{c|}{0} & \multicolumn{1}{c|}{1} & \multicolumn{1}{c|}{(-54.7316\textdegree, 32.5004\textdegree)} & 27 &  \multicolumn{1}{c|}{1} & \multicolumn{1}{c|}{0} & \multicolumn{1}{c|}{(-27.0498\textdegree, -84.5888\textdegree)} & \multicolumn{1}{c|}{} \\ \cline{1-8}

13 & \multicolumn{1}{c|}{0} & \multicolumn{1}{c|}{1} & \multicolumn{1}{c|}{(-59.5423\textdegree, -25.6850\textdegree)} & 28 & \multicolumn{1}{c|}{0} & \multicolumn{1}{c|}{1} & \multicolumn{1}{c|}{(-64.2131\textdegree, -9.1779\textdegree)} & \multicolumn{1}{c|}{} \\ \cline{1-8}

14 & \multicolumn{1}{c|}{0} & \multicolumn{1}{c|}{1} & \multicolumn{1}{c|}{(50.7947\textdegree, 5.1178\textdegree)} & 29 & \multicolumn{1}{c|}{0} & \multicolumn{1}{c|}{1} & \multicolumn{1}{c|}{(64.6667\textdegree, -29.3755\textdegree)} & \multicolumn{1}{c|}{} \\ \cline{1-8}

15 & \multicolumn{1}{c|}{0} & \multicolumn{1}{c|}{1} & \multicolumn{1}{c|}{(56.3838\textdegree, -17.5380\textdegree)} & 30 & \multicolumn{1}{c|}{1} & \multicolumn{1}{c|}{0} & \multicolumn{1}{c|}{(-29.4845\textdegree, 79.1200\textdegree)} & \multicolumn{1}{c|}{} \\ \hline

\end{tabular}%
}
\end{table*}

Fig.~\ref{fig: F_elements_vs_SEE} compares the Max-Min \gls{SEE} performance of the proposed \gls{RIS}-aided \gls{VLC} system for both the \gls{RSMA} and the power-domain \gls{NOMA} schemes for a different number of \gls{RIS} elements. In this figure, as the number of \gls{RIS} elements grows, we observe that the \gls{SEE} reaches a peak and then starts to decrease. By examining~\eqref{eq: SEE RSMA} or~\eqref{eq: SEE NOMA}, this trend can be explained by the logarithmic increase followed by the saturation behavior in \gls{SR}, the numerator, as illustrated in Fig.~\ref{fig: B_Elements_vs_SR_new}, and the linear increase in the total consumed energy, the denominator, with more \gls{RIS} elements. Practically, when the number of \gls{RIS} elements is high, the \gls{SR} improvement becomes unnoticeable, and the total consumed energy keeps increasing linearly. With such deployment, the system becomes energy-inefficient. Also, this figure shows that, for both scenarios, the best \gls{SEE} performance can be achieved when the mirror array-based \gls{RIS} contains two hundred elements. At this point, both schemes strike a favorable balance between the \gls{SR} improvement and the total energy consumption.

\begin{figure}[!t]
\centering
\vspace{-2em}
\includegraphics[width=0.4\textwidth]{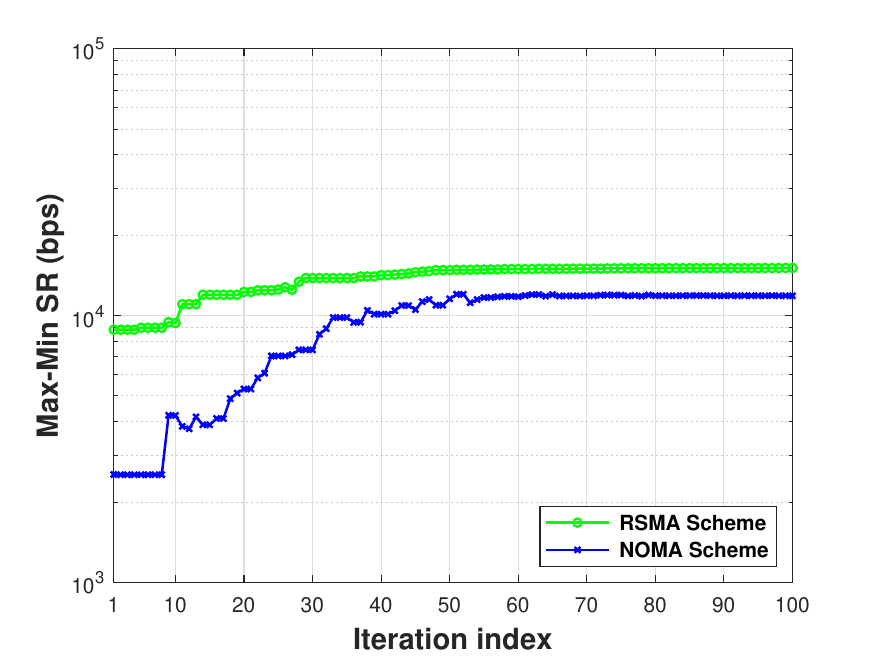}
\caption{The convergence curves of the considered \gls{GA} algorithm.}
\label{fig: G_Covergance_figure}
\vspace{-1em}
\end{figure}

\begin{figure}[!t]
\centering
\vspace{-2em}
\includegraphics[width=0.485\textwidth]{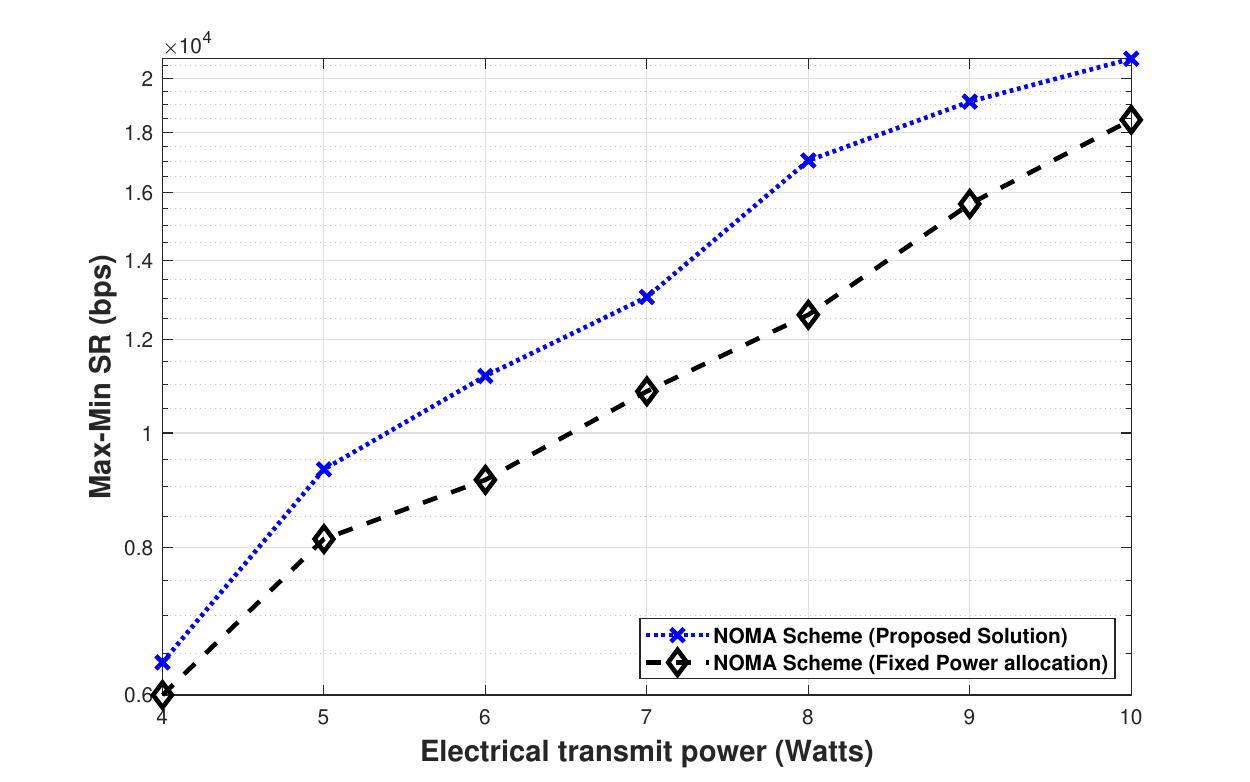}
\caption{Max-Min \gls{SR} performance versus \gls{VLC} \gls{AP} electrical transmit power for the proposed solution compared to the fixed power allocation baseline scheme.}
\label{fig: H_Baselines}
\vspace{-1em}
\end{figure}

\subsection{Convergence, Benchmarking, and Optimization Decision Variables Example} 
In this subsection, we present some results to provide deep insights into the adopted solution methodology and the optimized decision variables. For consistency, it is worth noting that the presented results are simulated while considering the following parameters: $P_S=5$ Watts, $K=30$, and $U=2$. Fig.~\ref{fig: G_Covergance_figure} shows the convergence curves of the \gls{GA} algorithm for the proposed mirror array-based \gls{RIS}-aided secure \gls{VLC} system for both the \gls{RSMA} and the power-domain \gls{NOMA} schemes. For both schemes, the \gls{GA} converges in less than $100$ iterations.

In Fig.~\ref{fig: H_Baselines}, our proposed power-domain \gls{NOMA}-based \gls{RIS}-aided \gls{VLC} system that jointly optimizes the \gls{VLC} \gls{AP} power allocation, \gls{RIS} association, and \gls{RIS} elements orientation angles is benchmarked with a counterpart system that uses a fixed power allocation scheme and jointly optimizes both the \gls{RIS} association and orientation using Algorithm~\ref{Algorithm: proposed solution 1}~\cite{7972998}. To achieve the fixed power allocation scheme, the parameter $\epsilon^{\textnormal{NOMA}}$ is set to $0.6$~\cite{10375270}. Subsequently, the \gls{VLC} \gls{AP} power allocation coefficient variables (i.e., $c_u$) are turned into constant values. This figure demonstrates that the proposed system achieves up to 135\% improvement in the Max-Min \gls{SR} performance compared to the counterpart scheme.

To better visualize the \gls{VLC} \gls{AP} power allocation, \gls{RIS} association, and \gls{RIS} elements orientation angles decision variables, an illustration is provided in Table~\ref{tab: optimized parameters}. Specifically, a concrete and compact example of the optimized decision variables for the \gls{RSMA} scheme is presented. As shown in the table, the entries of the \gls{RIS} association matrix $\textbf{G}$ are binary variables that take a value of either $0$ or $1$. The \gls{RIS} orientation angles variables ($\omega,\gamma$) lie within the range of $[-\frac{\pi}{2},\frac{\pi}{2}]$. It should be noted that negative angles cause reflections in a specific direction, while positive angles cause reflections in the opposite direction~\cite{10511290}. Lastly, the \gls{VLC} \gls{AP} power allocation variables ($P_0, P_1, P_2$) are continuous variables that fall within the range of ($0,1$).  In this example, given the aforementioned simulation parameters, the number of decision variables is $\mathcal{S}^\textnormal{RSMA}=(U*K)+(2*K)+(1+U)=(2*30)+(2*30)+(1+2)=123$. Also, as depicted in Table~\ref{tab: optimized parameters}, the number of elements allocated for User $1$ and User $2$ appears to be $6$ and $24$, respectively. Such \gls{RIS} element allocation achieves the minimum rate requirements of both users. Moreover, one can notice from the last column of Table~\ref{tab: optimized parameters} that most of the \gls{VLC} \gls{AP} transmit power is allocated for the common stream (i.e., $P_0$) compared to the allocated powers for the private streams (i.e., $P_1$ and $P_2$). This is expected and consistent with the \gls{SIC} requirements in \gls{RSMA}-based systems~\cite{9461768}.

\section{Conclusion and Future Research Directions}
\label{Sec: Conclusion}
This paper proposes and evaluates two optimization frameworks that maximize both the minimum \gls{SR} and the minimum \gls{SEE} for both power-domain \gls{NOMA} and \gls{RSMA}-based \gls{VLC} systems. Specifically, the proposed frameworks jointly consider the orientation angles of the \gls{RIS} elements, the association matrix, and the power allocation coefficients and then jointly solve the formulated problems through the \gls{GA}. To ensure efficient utilization of network resources, achieve high network spectral and energy efficiencies, and enhance user fairness, both the \gls{RSMA} and the power-domain \gls{NOMA} schemes are investigated. Also, our proposed model incorporates the random device orientation to extend the practicality and robustness of the considered \gls{RIS}-aided \gls{VLC} system. Comprehensive simulations are presented and revealed that the \gls{RSMA} scheme outperforms the power-domain \gls{NOMA} scheme in the proposed system, based on both the \gls{SR} and \gls{SEE} metrics and on various network parameters. In the future, we plan to (i) evaluate the proposed system in a larger room with multiple \gls{VLC} \glspl{AP}, (ii) analyze the impact of obstacles and (iii) extend the proposed framework to incorporate the \gls{PAPR} constraint. 

\bibliographystyle{IEEEtran}
\bibliography{main.bib}

\begin{thebibliography}{10}
\providecommand{\url}[1]{#1}
\csname url@samestyle\endcsname
\providecommand{\newblock}{\relax}
\providecommand{\bibinfo}[2]{#2}
\providecommand{\BIBentrySTDinterwordspacing}{\spaceskip=0pt\relax}
\providecommand{\BIBentryALTinterwordstretchfactor}{4}
\providecommand{\BIBentryALTinterwordspacing}{\spaceskip=\fontdimen2\font plus
\BIBentryALTinterwordstretchfactor\fontdimen3\font minus \fontdimen4\font\relax}
\providecommand{\BIBforeignlanguage}[2]{{%
\expandafter\ifx\csname l@#1\endcsname\relax
\typeout{** WARNING: IEEEtran.bst: No hyphenation pattern has been}%
\typeout{** loaded for the language `#1'. Using the pattern for}%
\typeout{** the default language instead.}%
\else
\language=\csname l@#1\endcsname
\fi
#2}}
\providecommand{\BIBdecl}{\relax}
\BIBdecl

\bibitem{7239528}
P.~H. Pathak, X.~Feng, P.~Hu, and P.~Mohapatra, ``Visible light communication, networking, and sensing: A survey, potential and challenges,'' \emph{IEEE Commun. Surv. Tutor.}, vol.~17, no.~4, pp. 2047--2077, 4th Quart. 2015.

\bibitem{9208801}
N.~Chi, Y.~Zhou, Y.~Wei, and F.~Hu, ``Visible light communication in {6G}: Advances, challenges, and prospects,'' \emph{IEEE Veh. Technol. Mag.}, vol.~15, no.~4, pp. 93--102, Sep. 2020.

\bibitem{8698841}
L.~E.~M. Matheus, A.~B. Vieira, L.~F.~M. Vieira, M.~A.~M. Vieira, and O.~Gnawali, ``Visible light communication: Concepts, applications and challenges,'' \emph{IEEE Commun. Surv. Tutor.}, vol.~21, no.~4, pp. 3204--3237, 4th Quart. 2019.

\bibitem{8528460}
S.~Al-Ahmadi, O.~Maraqa, M.~Uysal, and S.~M. Sait, ``Multi-user visible light communications: State-of-the-art and future directions,'' \emph{IEEE Access}, vol.~6, pp. 70\,555--70\,571, 2018.

\bibitem{10736549}
M.~H. Khoshafa, O.~Maraqa, J.~M. Moualeu, S.~Aboagye, T.~M.~N. Ngatched, M.~H. Ahmed, Y.~Gadallah, and M.~D. Renzo, ``{RIS}-assisted physical layer security in emerging {RF} and optical wireless communication systems: A comprehensive survey,'' \emph{IEEE Commun. Surv. Tutor.}, pp. 1--1, early access, 2024.

\bibitem{9968053}
S.~Aboagye, A.~R. Ndjiongue, T.~M.~N. Ngatched, O.~A. Dobre, and H.~V. Poor, ``{RIS}-assisted visible light communication systems: A tutorial,'' \emph{IEEE Commun. Surv. Tutor.}, vol.~25, no.~1, pp. 251--288, 1st Quart. 2023.

\bibitem{9744412}
Y.~Chen, Y.~Wang, J.~Zhang, P.~Zhang, and L.~Hanzo, ``Reconfigurable intelligent surface ({RIS})-aided vehicular networks: Their protocols, resource allocation, and performance,'' \emph{IEEE Veh. Technol. Mag.}, vol.~17, no.~2, pp. 26--36, Jun. 2022.

\bibitem{10158691}
Y.~Chen, Y.~Wang, X.~Guo, Z.~Han, and P.~Zhang, ``Location tracking for reconfigurable intelligent surfaces aided vehicle platoons: Diverse sparsities inspired approaches,'' \emph{IEEE J. Sel. Areas Commun.}, vol.~41, no.~8, pp. 2476--2496, Aug. 2023.

\bibitem{9984171}
Y.~Chen, Y.~Wang, and Z.~Wang, ``Reconfigurable intelligent surface aided high-mobility millimeter wave communications with dynamic dual-structured sparsity,'' \emph{IEEE Trans. Wireless Commun.}, vol.~22, no.~7, pp. 4580--4599, Jul. 2023.

\bibitem{9831440}
Y.~Mao, O.~Dizdar, B.~Clerckx, R.~Schober, P.~Popovski, and H.~V. Poor, ``Rate-splitting multiple access: Fundamentals, survey, and future research trends,'' \emph{IEEE Commun. Surv. Tutor.}, vol.~24, no.~4, pp. 2073--2126, 4th Quart. 2022.

\bibitem{10038476}
B.~Clerckx, Y.~Mao, E.~A. Jorswieck, J.~Yuan, D.~J. Love, E.~Erkip, and D.~Niyato, ``A primer on rate-splitting multiple access: Tutorial, myths, and frequently asked questions,'' \emph{IEEE J. Sel. Areas Commun.}, vol.~41, no.~5, pp. 1265--1308, May 2023.

\bibitem{10155552}
S.~Han, H.~Xia, X.~Zhou, and C.~Li, ``Securing {RSMA}-based communications at physical layer,'' \emph{IEEE Netw.}, vol.~38, no.~2, pp. 211--217, Mar. 2024.

\bibitem{9585491}
H.~Bastami, M.~Letafati, M.~Moradikia, A.~Abdelhadi, H.~Behroozi, and L.~Hanzo, ``On the physical layer security of the cooperative rate-splitting-aided downlink in {UAV} networks,'' \emph{IEEE Trans. Inf. Forensics Secur.}, vol.~16, pp. 5018--5033, Nov. 2021.

\bibitem{9154358}
O.~Maraqa, A.~S. Rajasekaran, S.~Al-Ahmadi, H.~Yanikomeroglu, and S.~M. Sait, ``A survey of rate-optimal power domain {NOMA} with enabling technologies of future wireless networks,'' \emph{IEEE Commun. Surv. Tutor.}, vol.~22, no.~4, pp. 2192--2235, 4th Quart. 2020.

\bibitem{10185552}
S.~Pakravan, J.-Y. Chouinard, X.~Li, M.~Zeng, W.~Hao, Q.-V. Pham, and O.~A. Dobre, ``Physical layer security for {NOMA} systems: Requirements, issues, and recommendations,'' \emph{IEEE Internet Things J.}, vol.~10, no.~24, pp. 21\,721--21\,737, Dec. 2023.

\bibitem{10598216}
N.~T. and B.~A. V., ``Full-duplex cooperative {NOMA} network with multiple eavesdroppers and non-ideal system imperfections: Analysis of physical layer security and validation using deep learning,'' \emph{IEEE Trans. Veh. Technol.}, vol.~73, no.~11, pp. 17\,192--17\,208, Nov. 2024.

\bibitem{9756553}
S.~Sun, F.~Yang, J.~Song, and Z.~Han, ``Optimization on multiuser physical layer security of intelligent reflecting surface-aided {VLC},'' \emph{IEEE Wirel. Commun. Lett.}, vol.~11, no.~7, pp. 1344--1348, Jul. 2022.

\bibitem{9784887}
D.~A. Saifaldeen, B.~S. Ciftler, M.~M. Abdallah, and K.~A. Qaraqe, ``{DRL}-based {IRS}-assisted secure visible light communications,'' \emph{IEEE Photonics J.}, vol.~14, no.~6, pp. 1--9, Dec. 2022.

\bibitem{10511290}
D.~A. Saifaldeen, A.~M. Al-Baseer, B.~S. Ciftler, M.~M. Abdallah, and K.~A. Qaraqe, ``{DRL}-based {IRS}-assisted secure hybrid visible light and {mmWave} communications,'' \emph{IEEE Open J. Commun. Soc.}, vol.~5, pp. 3007--3020, 2024.

\bibitem{10516681}
J.-Y. Wang, L.-H. Hong, N.~Liu, H.-N. Yang, P.~Feng, and J.~Ren, ``Secrecy analysis and optimization for {IRMA}-and jammer-aided visible light communications,'' \emph{IEEE Wirel. Commun. Lett.}, vol.~13, no.~7, pp. 1908--1912, Jul. 2024.

\bibitem{9500409}
L.~Qian, X.~Chi, L.~Zhao, and A.~Chaaban, ``Secure visible light communications via intelligent reflecting surfaces,'' in \emph{IEEE Int. Conf. Commun. (ICC), Montreal, QC, Canada}, Jun. 2021, pp. 1--6.

\bibitem{10669590}
R.~Iqbal, M.~Biagi, A.~Zoha, M.~A. Imran, and H.~Abumarshoud, ``Leveraging {IRS} induced time delay for enhanced physical layer security in {VLC} systems,'' \emph{IEEE Wirel. Commun. Lett.}, vol.~13, no.~11, pp. 3147--3151, Nov. 2024.

\bibitem{10279487}
H.~Abumarshoud, C.~Chen, I.~Tavakkolnia, H.~Haas, and M.~A. Imran, ``Intelligent reflecting surfaces for enhanced physical layer security in {NOMA} {VLC} systems,'' in \emph{IEEE Int. Conf. Commun. (ICC), Rome, Italy}, May 2023, pp. 3284--3289.

\bibitem{10746540}
Z.~Liu, F.~Yang, S.~Sun, J.~Song, and Z.~Han, ``Physical layer security in {NOMA}-based {VLC} systems with optical intelligent reflecting surface: A {Max-Min} secrecy data rate perspective,'' \emph{IEEE Internet Things J.}, vol.~12, no.~6, pp. 7180--7194, Mar. 2025.

\bibitem{7747504}
X.~Chen and M.~Jiang, ``Adaptive statistical bayesian {MMSE} channel estimation for visible light communication,'' \emph{IEEE Trans. Signal Process.}, vol.~65, no.~5, pp. 1287--1299, Mar. 2017.

\bibitem{9276478}
A.~M. Abdelhady, A.~K.~S. Salem, O.~Amin, B.~Shihada, and M.-S. Alouini, ``Visible light communications via intelligent reflecting surfaces: Metasurfaces vs mirror arrays,'' \emph{IEEE Open J. Commun. Soc.}, vol.~2, pp. 1--20, 2021.

\bibitem{8540452}
M.~D. Soltani, A.~A. Purwita, Z.~Zeng, H.~Haas, and M.~Safari, ``Modeling the random orientation of mobile devices: Measurement, analysis and {LiFi} use case,'' \emph{IEEE Trans. Commun.}, vol.~67, no.~3, pp. 2157--2172, Mar. 2019.

\bibitem{9543660}
S.~Aboagye, T.~M.~N. Ngatched, O.~A. Dobre, and A.~R. Ndjiongue, ``Intelligent reflecting surface-aided indoor visible light communication systems,'' \emph{IEEE Commun. Lett.}, vol.~25, no.~12, pp. 3913--3917, Dec. 2021.

\bibitem{9464264}
S.~Cho, G.~Chen, and J.~P. Coon, ``Cooperative beamforming and jamming for secure {VLC} system in the presence of active and passive eavesdroppers,'' \emph{IEEE Trans. Green Commun. Netw.}, vol.~5, no.~4, pp. 1988--1998, Dec. 2021.

\bibitem{9270035}
F.~Rottenberg, T.-H. Nguyen, J.-M. Dricot, F.~Horlin, and J.~Louveaux, ``{CSI}-based versus {RSS}-based secret-key generation under correlated eavesdropping,'' \emph{IEEE Trans. Commun.}, vol.~69, no.~3, pp. 1868--1881, Mar. 2021.

\bibitem{maraqa2021achievable}
O.~Maraqa, U.~F. Siddiqi, S.~Al-Ahmadi, and S.~M. Sait, ``On the achievable {Max-Min} user rates in multi-carrier centralized {NOMA-VLC} networks,'' \emph{Sensors}, vol.~21, no.~11, p. 3705, May 2021.

\bibitem{9032167}
K.~Cao, B.~Wang, H.~Ding, T.~Li, J.~Tian, and F.~Gong, ``Secure transmission designs for {NOMA} systems against internal and external eavesdropping,'' \emph{IEEE Trans. Inf. Forensics Secur.}, vol.~15, pp. 2930--2943, 2020.

\bibitem{9780419}
M.~Abolpour, S.~Aïssa, L.~Musavian, and A.~Bhowal, ``Rate splitting in the presence of untrusted users: Outage and secrecy outage performances,'' \emph{IEEE Open J. Commun. Soc.}, vol.~3, pp. 921--935, 2022.

\bibitem{10375270}
O.~Maraqa, S.~Aboagye, and T.~M.~N. Ngatched, ``Optical {STAR}-{RIS}-aided {VLC} systems: {RSMA} versus {NOMA},'' \emph{IEEE Open J. Commun. Soc.}, vol.~5, pp. 430--441, 2024.

\bibitem{wang2016physical}
H.-M. Wang and T.-X. Zheng, \emph{{Physical Layer Security in Random Cellular Networks}}.\hskip 1em plus 0.5em minus 0.4em\relax Singapore: Springer, 2016.

\bibitem{9195771}
H.~Fu, S.~Feng, W.~Tang, and D.~W.~K. Ng, ``Robust secure beamforming design for two-user downlink {MISO} rate-splitting systems,'' \emph{IEEE Trans. Wirel. Commun.}, vol.~19, no.~12, pp. 8351--8365, Dec. 2020.

\bibitem{shen2023secrecy}
T.~Shen, V.~Yachongka, Y.~Hama, and H.~Ochiai, ``Secrecy design of indoor visible light communication network under downlink {NOMA} transmission,'' \emph{arXiv preprint arXiv:2304.08458}, Apr. 2023.

\bibitem{8713985}
C.~Du, F.~Zhang, S.~Ma, Y.~Tang, H.~Li, H.~Wang, and S.~Li, ``Secure transmission for downlink {NOMA} visible light communication networks,'' \emph{IEEE Access}, vol.~7, pp. 65\,332--65\,341, 2019.

\bibitem{Sait:1999}
S.~M. Sait and H.~Youssef, \emph{{Iterative Computer Algorithms with Applications in Engineering: Solving Combinatorial Optimization Problems}}.\hskip 1em plus 0.5em minus 0.4em\relax Los Alamitos, CA, USA: IEEE Computer Society Press, 1999.

\bibitem{10445725}
H.~Wang, Z.~Han, and A.~L. Swindlehurst, ``Channel reciprocity attacks using intelligent surfaces with non-diagonal phase shifts,'' \emph{IEEE Open J. Commun. Soc.}, vol.~5, pp. 1469--1485, 2024.

\bibitem{10500850}
S.~Abdeljabar, M.~W. Eltokhey, and M.-S. Alouini, ``Sum rate and fairness optimization in {RIS}-assisted {VLC} systems,'' \emph{IEEE Open J. Commun. Soc.}, vol.~5, pp. 2555--2566, 2024.

\bibitem{9463422}
S.~T. Duong, T.~V. Pham, C.~T. Nguyen, and A.~T. Pham, ``Energy-efficient precoding designs for multi-user visible light communication systems with confidential messages,'' \emph{IEEE Trans. Green Commun. Netw.}, vol.~5, no.~4, pp. 1974--1987, Dec. 2021.

\bibitem{9633190}
D.~R. Pattanayak, V.~K. Dwivedi, V.~Karwal, A.~Upadhya, H.~Lei, and G.~Singh, ``Secure transmission for energy efficient parallel mixed {FSO/RF} system in presence of independent eavesdroppers,'' \emph{IEEE Photonics J.}, vol.~14, no.~1, pp. 1--14, Feb. 2022.

\bibitem{10404069}
T.~V. Pham, A.~T. Pham, and S.~Ishihara, ``Design of energy-efficient artificial noise for physical layer security in visible light communications,'' \emph{IEEE Trans. Green Commun. Netw.}, vol.~8, no.~2, pp. 741--755, Jun. 2024.

\bibitem{8307185}
S.~Ma, T.~Zhang, S.~Lu, H.~Li, Z.~Wu, and S.~Li, ``Energy efficiency of {SISO} and {MISO} in visible light communication systems,'' \emph{J. Light. Technol.}, vol.~36, no.~12, pp. 2499--2509, Jun. 2018.

\bibitem{mitchell1998}
M.~Mitchell, \emph{{An Introduction to Genetic Algorithms}}.\hskip 1em plus 0.5em minus 0.4em\relax MIT press, 1998.

\bibitem{8669970}
S.~Feng, T.~Bai, and L.~Hanzo, ``Joint power allocation for the multi-user {NOMA}-downlink in a power-line-fed {VLC} network,'' \emph{IEEE Trans. Veh. Tech.}, vol.~68, no.~5, pp. 5185--5190, May 2019.

\bibitem{10183987}
O.~Maraqa and T.~M.~N. Ngatched, ``Optimized design of joint mirror array and liquid crystal-based {RIS}-aided {VLC} systems,'' \emph{IEEE Photon. J.}, vol.~15, no.~4, pp. 1--11, Jul. 2023.

\bibitem{10959088}
R.~Ahiaklo-Kuz, S.~Aboagye, O.~Maraqa, and T.~M.~N. Ngatched, ``Design and optimization of an integrated visible light communication and localization system using liquid crystal based-{RIS} receivers,'' \emph{IEEE Photonics J.}, vol.~17, no.~3, pp. 1--9, Jun. 2025.

\bibitem{rasti2021design}
A.~Rasti-Meymandi, A.~Madahian, J.~Abouei, A.~Mirvakili, Z.~HajiAkhondi-Meybodi, A.~Mohammadi, and M.~Uysal, ``Design and implementation of {VLC}-based smart barrier gate systems,'' \emph{AEU-International Journal of Electronics and Communications}, vol. 136, p. 153765, Jul. 2021.

\bibitem{de2019batch}
V.~V. De~Melo, D.~V. Vargas, and W.~Banzhaf, ``Batch tournament selection for genetic programming: the quality of lexicase, the speed of tournament,'' in \emph{Proc. Genet. Evol. Comput. Conf. (GECCO), Prague, Czech Republic}, Jul. 2019, pp. 994--1002.

\bibitem{8491100}
Y.~Mao, B.~Clerckx, and V.~O. Li, ``Energy efficiency of rate-splitting multiple access, and performance benefits over {SDMA} and {NOMA},'' in \emph{Int. Symp. Wirel. Commun. Syst. (ISWCS), Lisbon, Portugal}, Aug. 2018, pp. 1--5.

\bibitem{7972998}
H.~Marshoud, P.~C. Sofotasios, S.~Muhaidat, G.~K. Karagiannidis, and B.~S. Sharif, ``On the performance of visible light communication systems with non-orthogonal multiple access,'' \emph{IEEE Trans. Wirel. Commun.}, vol.~16, no.~10, pp. 6350--6364, Oct. 2017.

\bibitem{9461768}
Z.~Yang, M.~Chen, W.~Saad, and M.~Shikh-Bahaei, ``Optimization of rate allocation and power control for rate splitting multiple access ({RSMA}),'' \emph{IEEE Trans. Commun.}, vol.~69, no.~9, pp. 5988--6002, Sep. 2021.

\end{thebibliography}

\vskip -2\baselineskip plus -1fil

\begin{IEEEbiography}[{\includegraphics[width=1in,height=1.25in, clip,keepaspectratio]{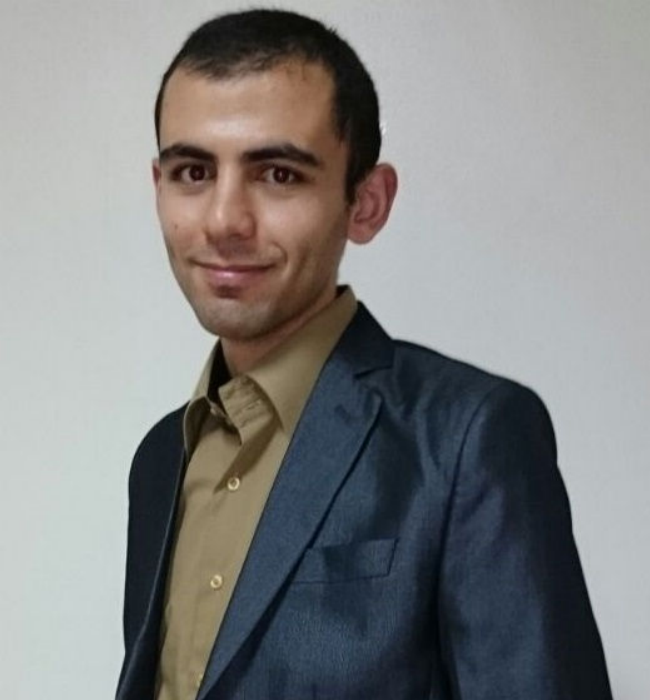}}]{\textbf{Omar Maraqa}} \hspace{0.01em} received the B.Eng. degree in Electrical Engineering from Palestine Polytechnic University, Palestine, in 2011, the M.Sc. degree in Computer Engineering and the Ph.D. degree in Electrical Engineering from King Fahd University of Petroleum \& Minerals (KFUPM), Dhahran, Saudi Arabia, in 2017 and 2022, respectively. He is currently a Postdoctoral Research Fellow with the Department of Electrical and Computer Engineering, at McMaster University, Canada. He was recognized as an exemplary reviewer for \textsc{IEEE Communications Letters} in 2023. He serves as a Technical Program Committee (TPC) Member for the IEEE Vehicular Technology Conference, and a reviewer for several IEEE journals. His research interests include optimization and performance analysis of emerging wireless communications systems.
\end{IEEEbiography}

\vskip -1\baselineskip plus -0fil

\begin{IEEEbiography}[{\includegraphics[width=1in,height=1.250in, clip,keepaspectratio]{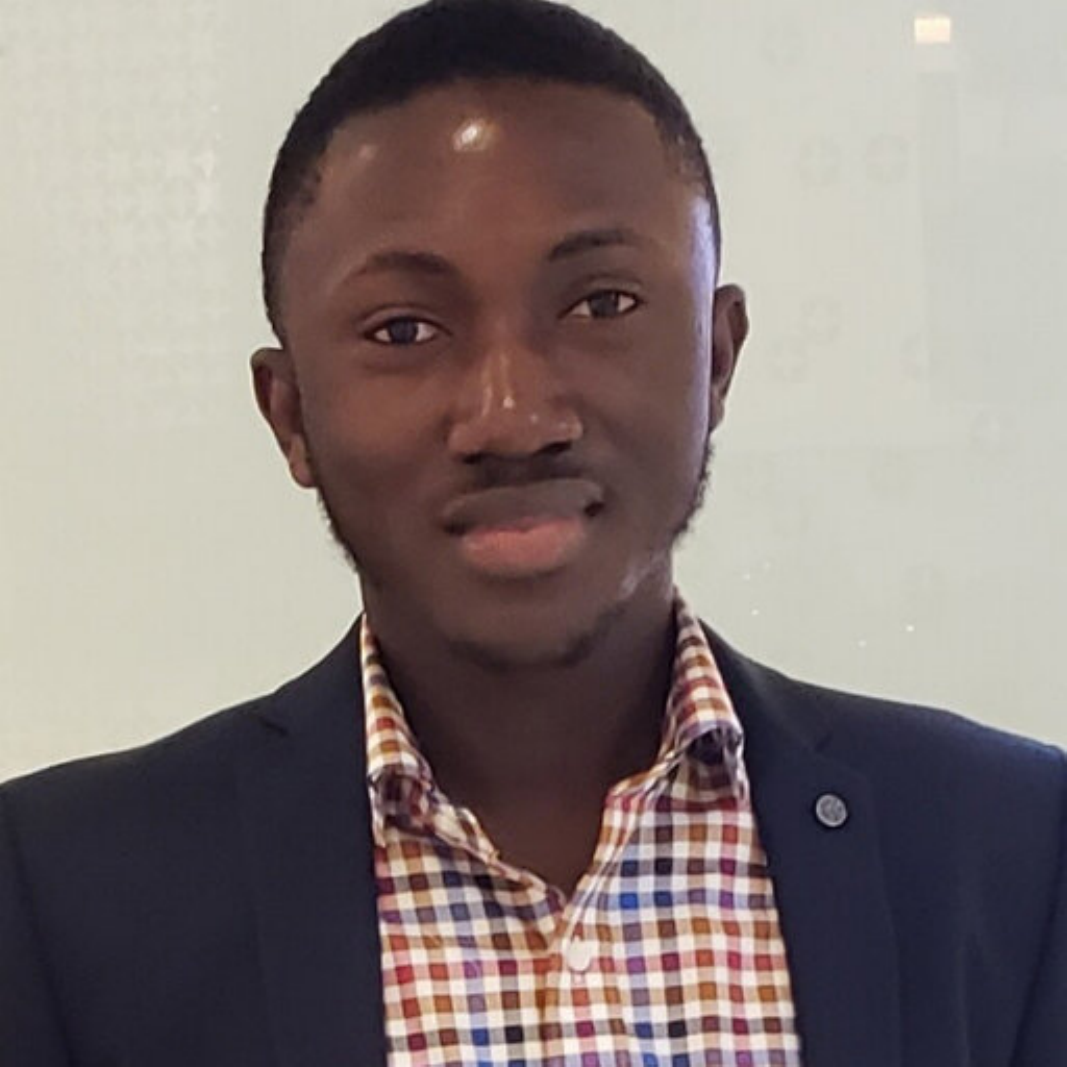}}]{\textbf{Sylvester Aboagye}} \hspace{0.01em} (Member, IEEE) received the B.Sc. degree (Hons.) in telecommunication engineering from the Kwame Nkrumah University of Science and Technology, Kumasi, Ghana, in 2015, and the M.Eng. and Ph.D. degrees in electrical engineering from Memorial University, St. John's, NL, Canada, in 2018 and 2022, respectively. He received many prestigious awards, including the Governor General's Academic Gold Medal. He was a Postdoctoral Research Fellow with the Department of Electrical Engineering and Computer Science at York University, Canada, from January to December 2023, and is currently an Assistant Professor with the School of Engineering, University of Guelph, Canada. His current research interests include the design and optimization of multiband wireless networks, visible light communication systems, terrestrial and non-terrestrial integrated sensing and communication networks, and 6G and beyond enabling technologies. Dr. Aboagye serves as an Editor for \textsc{IEEE Communications Letters} and \textsc{IEEE Open Journal of the Communications Society}, a Technical Program Committee (TPC) member for the IEEE Vehicular Technology Conference, and a reviewer for several IEEE journals. He was recognized as an Exemplary Reviewer by \textsc{IEEE Communications Letters} and \textsc{IEEE Transactions on Network Science and Engineering} in 2023.
\end{IEEEbiography}

\vskip -1\baselineskip plus 0fil

\begin{IEEEbiography}[{\includegraphics[width=1in,height=1.25in, clip,keepaspectratio]{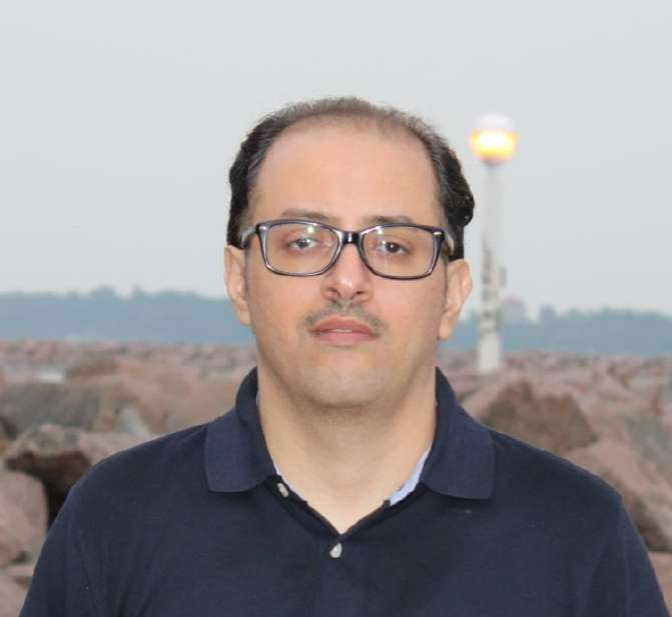}}]{\textbf{Majid H. Khoshafa}} \hspace{0.01em} (Senior Member, IEEE) received his B.Sc. in Communication Engineering from Ibb University, Yemen, in 2007, his M.Sc. in Telecommunications Engineering from King Fahd University of Petroleum and Minerals (KFUPM), Saudi Arabia, in 2017, and his Ph.D. in Electrical and Computer Engineering from Memorial University of Newfoundland (MUN), Canada, in 2022. From 2009 to 2010, he worked as a Radio Network Planning and Optimization Engineer at MTN Telecommunications, Sana'a, Yemen. From 2010 to 2013, he worked as a Lecturer Assistant in the Electrical and Computer Department, Faculty of Engineering, Ibb University, Ibb, Yemen. He is currently a Postdoctoral Fellow in the Department of Electrical and Computer Engineering at McMaster University, Canada. Previously, he was a Postdoctoral Research Fellow at Queen's University, Canada, from 2022 to 2023. His research interests include wireless communications, reconfigurable intelligent surface (RIS) technology, physical layer security, AI, and 6G and beyond enabling technologies. He has received several awards from MUN, including the Fellowship of the School of Graduate Studies, the Recognition of Excellence Award, the International Student Recognition Award, and the Graduate Academic Excellence Award. He serves as a Technical Program Committee (TPC) Member for the IEEE Vehicular Technology Conference and a Reviewer for several IEEE journals. He was recognized as an Exemplary Reviewer of \textsc{IEEE Wireless Communications Letters} and  \textsc{IEEE Communications Letters} in 2024. 
\end{IEEEbiography}

\vskip -1\baselineskip plus 0fil

\begin{IEEEbiography}[{\includegraphics[width=1in,height=1.25in, clip,keepaspectratio]{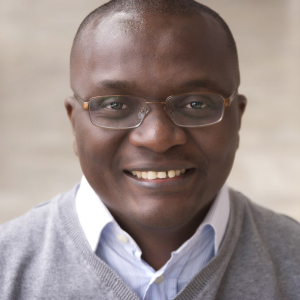}}]{\textbf{Telex M. N. Ngatched}} \hspace{0.01em} (Senior Member, IEEE) received the B.Sc. degree and the M.Sc. degree in electronics from the University of Yaound\'e, Cameroon, in 1992 and 1993, respectively, the MscEng (Cum Laude) in electronic engineering from the University of Natal, Durban, South Africa, in 2002, and the Ph.D. in electronic engineering from the University of KwaZulu-Natal, Durban, South Africa, in 2006. From July 2006 to December 2007, he was with the University of KwaZulu-Natal as a Postdoctoral Fellow, from 2008 to 2012 with the Department of Electrical and Computer Engineering, University of Manitoba, Canada, as a Research Associate, and from 2012 to 2022 with Memorial University. He joined McMaster University in January 2023, where he is currently an Associate Professor. His research interests include 5G and 6G enabling technologies, optical wireless communications, hybrid optical wireless and radio frequency communications, artificial intelligence and machine learning for communications, and underwater communications. Dr. Ngatched serves as an Area Editor for the \textsc{IEEE Open Journal of the Communications Society}, an Associate Technical Editor for the \textsc{IEEE Communications Magazine}, and an Editor of the \textsc{IEEE Communications Society On-Line Content}. He was a recipient of the Best Paper Award at the IEEE Wireless Communications and Networking Conference (WCNC) in 2019. He is a Professional Engineer (P. Eng.) registered with the Professional Engineers Ontario, Toronto, ON, Canada.
\end{IEEEbiography}

\end{document}